%% file: paper.tex
\documentclass[preprint,5p, twocolumn, authoryear]{elsarticle}

\usepackage[dvipsnames]{xcolor}
\usepackage{graphicx}
\usepackage{makecell}
\usepackage{multirow}
\usepackage{amsmath}   
\usepackage{amssymb} 
\usepackage{hyperref}
\hypersetup{colorlinks=false,
            pdfborder={0 0 0}}
\usepackage{amsthm}    
\usepackage{pifont}
\usepackage{units}
\usepackage{array}
\usepackage{booktabs}
\usepackage{placeins}
\usepackage{tabularx}
\newcolumntype{Y}{>{\centering\arraybackslash}X}
\usepackage{cleveref}

\crefname{appendix}{appendix}{appendices}
\Crefname{appendix}{Appendix}{Appendices}
\usepackage{xr}
\usepackage{enumitem}
\usepackage{tikz, tikzsettings}
\usepackage{mathtools}
\usepackage{mpcsymbols}
\usepackage[version=4]{mhchem} 
\usepackage[singlelinecheck=false]{caption}

\newcommand{\Rm}{\mathrm}
\DeclareMathOperator*{\argmin}{argmin}

\journal{Computers and Chemical Engineering}

\begin{document}

\begin{frontmatter}

\title{A tale of perfect fit and phantom optima:\\
how data-driven models can fail in real-time optimization}

\author[label1]{Prithvi Dake\corref{cor1}}
\ead{prithvidake@ucsb.edu}
\cortext[cor1]{Corresponding Author}
\author[label2]{Rahul Bindlish}
\ead{rbindlish@dow.com}
\author[label1]{James B. Rawlings}
\ead{jbraw@ucsb.edu}

\address[label1]{Department of Chemical Engineering, University of California,
          Santa Barbara, CA 93106, United States}
\address[label2]{Dow Chemical Company, TX, United States}

\begin{abstract}
Real-time optimization (RTO) relies on process models to locate economically 
optimal operating conditions. Because developing first-principles models requires 
significant process knowledge, data-driven alternatives are increasingly attractive. 
Modern machine-learning models can fit historical plant data accurately and often 
pass standard validation tests. Whether such models can be trusted for economic 
optimization, however, remains unclear.
We investigate this question using a vinyl acetate monomer benchmark process with
a unique, well-conditioned economic optimum. We train a structured hybrid model 
that combines known mass balances and thermodynamics with a neural-network closure 
for unknown kinetics, and a fully data-driven neural ordinary differential equation 
(ODE) model. Both models 
reproduce plant measurements accurately and exhibit little variation in predictions 
across random initializations. Yet their economic optima differ substantially 
from that of the plant. Where the plant returns a single optimum on multistart search, 
the trained models return many \emph{phantom} optima. 
We further show that the training optimizer alone can be yet another source of error. 
Even with noise-free data and initialization at weights that recover the plant 
optimum, stochastic gradient training can drift to weights that yield 
substantially worse RTO solutions. The identified model is thus an artifact of
the training optimizer as well as the data. 
These results demonstrate that a good predictive fit of all available measurements 
does not guarantee reliable economic performance. 
A data-driven model for RTO should at least be required to recover the optimum 
on a decision-oriented benchmark like the one developed here before being 
considered for plant testing and application.
\end{abstract}

\begin{keyword}
Real-time optimization \sep
Physics-constrained machine learning \sep
Neural ordinary differential equations \sep
Parameter estimation \sep
Decision-oriented benchmark
\end{keyword}

\end{frontmatter}

\section{Introduction}

Growing global competition and sustainability demands have made RTO
increasingly important to the chemical process industries
\citep{naysmith:douglas:1995,darby:nikolaou:jones:2011,camara:quelhos:pinto:2016}.
RTO operates on a timescale of hours to days, or is triggered
when a price change or disturbance is detected. It
supplies the profitable steady-state setpoints to lower-level regulatory layers
\citep{cutler:perry:lu:1983}.
Traditional RTO strategies often use a rigorous \emph{nonlinear steady-state}
process model that is updated using plant measurements
after each optimization iteration \citep[pg. 351]{seborg:edgar:mellichamp:doyle:2017}.

Specifically, the model parameters are reconciled against the most recent
steady-state plant measurements, a step termed \emph{model adaptation} 
or \emph{data reconciliation} \citep{crowe:1996}.
The updated model is then used to solve a steady-state economic optimization problem.
This two-step approach \citep{roberts:williams:1981} is a common way of practicing
RTO. While the plant tracks the RTO setpoint, it is in transient operation, 
and one must detect the new steady state before reconciling with measurements.
Thus, the approach is limited
by steady-state \emph{wait-times} and fragile detection algorithms
that require careful tuning \citep{darby:nikolaou:jones:2011}.

Recent developments in the two-step RTO approach adapt the steady-state
model using transient plant measurements \citep{krishnamoorthy:skogestad:2018}, 
which eliminates both the
steady-state wait-times and detection. The RTO problem can also be made dynamic, e.g.,
economic MPC \citep{rawlings:angeli:bates:2012}. However, when the disturbances vary slowly
relative to the plant's settling time,
steady-state RTO is preferred because it is cheaper to solve.

Besides model-based methods, model-free methods also exist.
RTO has been carried out using \emph{direct-search} methods that usually mimic
gradient-free schemes like Nelder-Mead or evolutionary strategies \citep{box:1957}.
Without gradient information, the number of plant evaluations grows steeply
with the decision space for these methods.

A second family of model-free methods can be understood
as a feedback problem in the form of extremum seeking or self-optimizing type control
\citep{skogestad:2000b}.
If we treat the gradient of profit with respect to the degrees of freedom 
at steady-state as a controlled variable and hold it at zero, the plant 
\emph{self-optimizes} to a stationary point of the RTO problem.
Most of the feedback methods therefore focus on \emph{gradient estimation}, 
e.g., using dither signals.
The feedback approach does not distinguish among a \emph{maximum}, \emph{minimum}, or
\emph{saddle point}, and
leads to prohibitively slow convergence and a difficult dither tuning problem
in practice \citep{krishnamoorthy:skogestad:2022}. For these reasons, 
model-free methods are not considered further.

A third method, \emph{modifier adaptation}, combines the two: it converges in
few plant iterations, 
like a model-based method, yet provably reaches the plant's optimum, like the 
feedback-based schemes \citep{bonvin:pannocchia:2024}.
Crucially, it recovers the plant optimum despite model mismatch, provided the 
model already has the correct curvature at the plant optimum 
(a positive semidefinite Hessian in the case of a minimum), so that 
the plant's gradient corrections suffice 
\citep{chachuat:srinivasan:bonvin:2009, papasavvas:ferreira:marchetti:bonvin:2019}. 
Yet this curvature is not guaranteed even for a model with 
no mismatch, as we show in this paper.
The accurate estimation of plant gradients remains another challenge.
We therefore focus on model-based RTO using transient measurements.

\begin{figure*}[!tp]
    \centering
    {\bfseries \Large
        \resizebox{0.65\textwidth}{!}{\input{abstract.tex}}}
    \caption{(A) Historical plant datasets are 
    used to train (B) a structured model and a black-box ML/AI closure model, 
    each deployed in real-time optimization (RTO). (C) Near-identical fits map 
    to very different optima, the worst-case structured model
    loses close to 30\% profit and the black-box model over 50\%.
    A \emph{reliable} data-driven model must recover the optimum, not be judged solely
    on the validation dataset.}
    \label{fig:abstract}
\end{figure*}
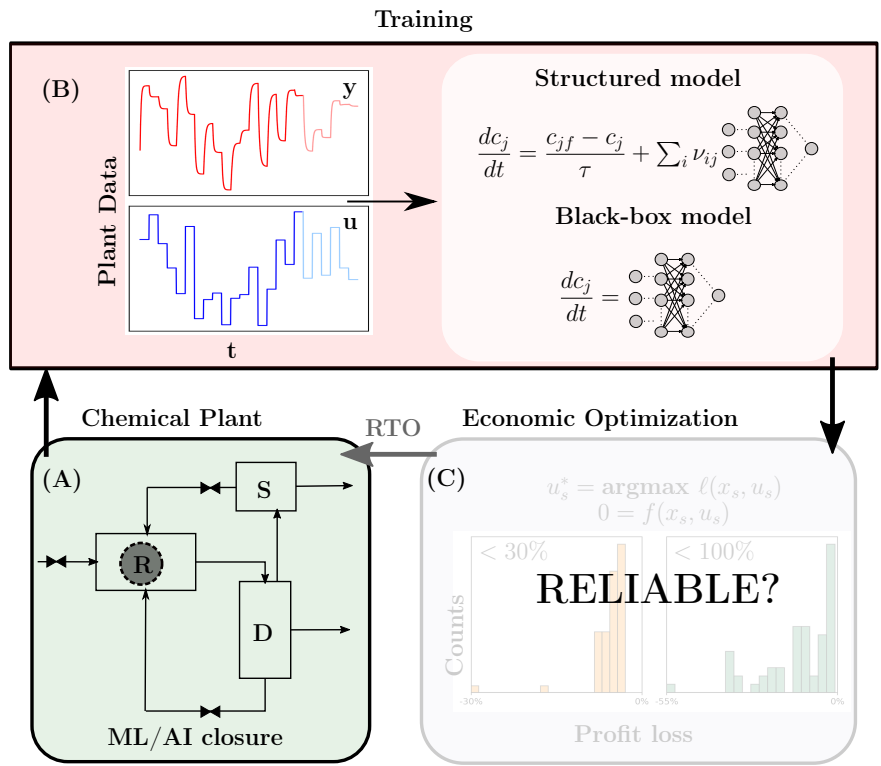

\subsection{Modeling}

For RTO, practitioners usually start with first-principles models (FPMs), built on
conservation laws and thermodynamics. The model's structure leaves few free
parameters, so it extrapolates well beyond 
the operating data 
and remains interpretable. These are exactly the
properties an economic optimizer relies upon \citep{bindlish:2025}. 
But the cost to build and maintain such models 
is large since FPMs demand more
process knowledge than is often available. 

An FPM is only as good as its assumed structure. When that structure is 
incomplete, e.g., when the reaction kinetics or fluid dynamics are not 
known from first principles, the model violates the \emph{adequacy} 
condition of \citet{forbes:marlin:macgregor:1994}: no choice of parameters 
satisfies the plant's optimality conditions, and its economic optimum is 
biased. 

Alternatives include purely data-driven or \emph{black-box} models that learn plant dynamics
from input--output data with little structure imposed. Such models 
use neural networks (NNs) that are universal function approximators 
\citep{hornik:stinchcombe:white:1989}. 
Neural ordinary differential
equations \citep{chen:rubanova:bettencourt:duvenaud:2018} are a
representative modern instance of such black-box models. 
They need little process
knowledge, but they do not preserve conservation laws and degrade rapidly outside the
training set. Violating conservation laws leads to infeasible predictions, 
which is usually the biggest concern in 
chemical engineering applications.

Augmenting the first-principles balances with a data-driven closure for the 
unknown parts in FPMs has therefore been a long-standing pursuit in chemical 
engineering (see the reviews by 
\citealp{sansana:joswiak:castillo:wang:rendall:chiang:reis:2021,
schweidtmann:zhang:stosch:2024}).
The resulting modeling techniques have been recently surveyed under the umbrella of \emph{physics-constrained
machine learning} (PCML) by \citet{mukherjee:zavala:2026}. PCML strategies are 
commonly grouped by how the physics is enforced. In \emph{soft}-constrained 
models, the known physical laws enter the loss as a penalty, as in physics-informed 
neural networks \citep{raissi:perdikaris:karniadakis:2019, sholokhov:liu:mansour:nabi:2023}. 
In contrast, \emph{hard}-constrained models enforce steady-state mass and energy balances exactly 
\citep{golder:roy:hasan:2025, constante:chen:li:2026}.
Another example includes the \emph{simultaneous projection} method, which is seen as a promising step 
toward using NNs to approximate parts of DAE models 
\citep{lueg:alves:schicksnus:kitchin:laird:biegler:2025}. Finally, in \emph{hybrid} 
(or \emph{structured}) models, a data-driven term is wired into a partially 
mechanistic transient model 
and solved with \emph{shooting} methods \citep{kumar:rawlings:2023a, thompson:connors:zavala:venturelli:2026}. 
The physics-based dynamic balances are 
retained, while a data-learned term stands in for the part that is not known from 
first principles, typically the reaction kinetics or fluid dynamics. We use 
both hybrid and black-box models in this work.

Recent perspectives highlight the absence 
of decision-oriented benchmarks as central open problems in PCML 
\citep{pantelides:baldea:georgiou:gopaluni:mehmet:sheth:zavala:georgakis:2025,
mukherjee:zavala:2026}. At the same time, 
a growing literature argues that reliable models for process systems can be 
built largely from historical plant data using machine learning (ML) and 
artificial intelligence (AI) \citep{sansana:joswiak:castillo:wang:rendall:chiang:reis:2021,
schweidtmann:zhang:stosch:2024}. We take up both threads and ask: for a 
data-driven model built with the best modeling paradigms and software available to us, 
carrying as much physical structure as the process permits, and clearing 
the usual validation checks, does the optimum it yields under RTO match 
the plant's?
A graphical abstract of the work is shown in \cref{fig:abstract}. 
Our contributions are as follows:

\begin{enumerate}[label=(\roman*)]
    \item Evidence that near-identical fits, tight even across random 
    initializations, still yield sharply disparate economic optima, so standard 
    validation and uncertainty quantification do not certify reliable 
    RTO performance.
    \item A controlled demonstration that the \emph{stochastic} training optimizer alone can cause
    divergence from the optimal weights. Given correct model structure, noise-free data, and
    initialization at weights at which the model recovers the plant optimum, stochastic
    gradient training still drifts to a model whose RTO solution is significantly suboptimal.
    \item A decision-oriented benchmark, the vinyl acetate process with a
    well-conditioned economic optimum, on which a data-driven RTO model must
    recover the plant optimum rather than merely fit plant data.
\end{enumerate}

\textit{Outline} The paper is organized as follows: \Cref{sec:casestudy} introduces the
vinyl acetate process used as the case study, sets up its economic optimization,
and describes the structured (hybrid) and black-box (neural ODE) models.
\Cref{sec:training} gives the training data, scaling, initialization, and
software details. \Cref{sec:results} reports the results, comparing the
parametric, structured, and black-box models on how well they recover the plant
economic optimum. \Cref{sec:conclusions} concludes.
Further details on the flowsheet are provided in \Cref{app:process_units,app:parameter_values}.
The supplementary material reports the hyperparameter tuning and the 
training optimizer comparison for every case studied in this paper.

\section{Industrial case-study: Vinyl acetate process}
\label{sec:casestudy}

The vinyl acetate monomer (VAc) production process has been used as a benchmark problem
for plantwide control study due to its challenging and coupled process network 
\citep{luyben:tyreus:1998, chen:dave:mcavoy:luyben:2003}. 

We model the process using a detailed reactor model coupled with a simplified separator
system, which includes the absorber, flash drum and distillation column.
Although the separator system is simplified, the nonlinear reactor model provides
rich insight into the process for steady-state economic optimization \citep{ward:mellichamp:doherty:2004}. 
\subsection{Process description}
\label{sec:VAc_model}

\begin{figure*}[!tp]
    \centering
    {\bfseries \Large
        \resizebox{0.7\textwidth}{!}{\input{Economics.tex}}}
    \caption{Simplified process flowsheet for the vinyl acetate (VAc) monomer 
    plant consisting of 7 units and 13 streams. (Ee: Ethylene, Ea: Ethane, A: 
    Acetic acid, O: Oxygen, W: Water, V: Vinyl acetate, C: Carbon dioxide). 
    Outlet concentrations from the reactor are measured subject to step inputs
    in the decision variables (fresh feed rates, recycle fractions and reactor temperature),
    that are shown by the control valves and the controller icon.}
    \label{fig:block_measurement}
\end{figure*}
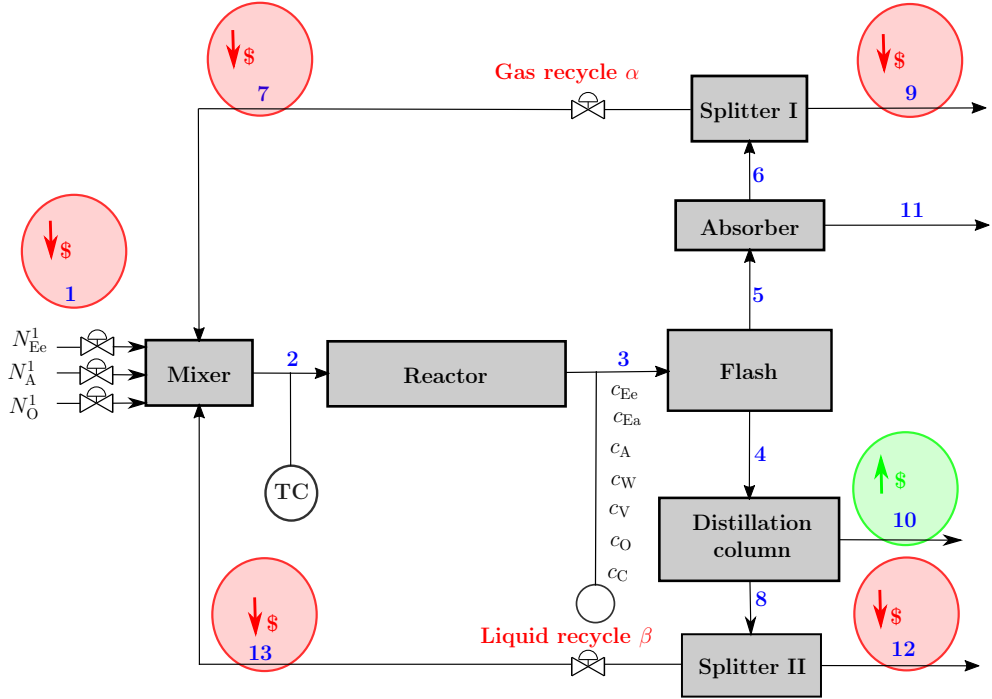

The VAc process involves seven chemical species: ethylene (\ce{C2H4}) with inert ethane (\ce{C2H6}), 
oxygen (\ce{O2}) and acetic acid (\ce{CH3COOH}) as fresh feeds; vinyl acetate (\ce{C4H6O2}) 
is the product; water (\ce{H2O}) and carbon dioxide (\ce{CO2}) are byproducts. 
The following reactions take place inside the reactor:
{\small
\begin{align}
    \ce{C2H4 + CH3COOH + 1/2 O2 &->[r_1] C4H6O2 + H2O} \notag \\
    \ce{C2H4 + 3 O2 &->[r_2] 2 CO2 + 2 H2O}
    \label{eq:rxn}
\end{align}
}
We develop a simplified flowsheet as shown in \cref{fig:block_measurement}. In the
process, the chemical species (in gaseous phase) coming from the mixer (stream 2)
are fed to the reactor. The gaseous effluent from the reactor (stream 3) is partly
condensed to separate into gas and liquid in the flash drum. The gases
(stream 5) then pass through a \ce{CO2} absorber, and stream 6 is then partly 
recycled back to the mixer by the splitter I (stream 7). The liquid stream 4 is
then sent to a distillation column. The overhead products (mainly vinyl acetate
and water) are withdrawn (stream 10) while the bottom products (mainly acetic
acid and water) are recycled back to the mixer by splitter II (stream 13). Thus,
the flowsheet constitutes a conventional reactor-separator system with two recycles.
Note that the reactions in \cref{eq:rxn} are not elementary but rather an overall
representation of the actual reaction network. We also assume a lumped model of the catalyst,
such that all the mass transfer resistances are represented
by the following rate laws:
\begin{align}
    r_1 &=  k_1 \times 
            e^{-E_1/RT} \times 
            c_\Rm{Ee}^{1.1} c_\Rm{O}^{1.1}
            c_\Rm{A} \notag \\
    r_2 &=  k_2 \times
            e^{-E_2/RT} \times 
            c_\Rm{Ee} c_\Rm{O} 
            \label{eq:rxnrate}
\end{align}
where $c_j, T$ are the concentration of species $j$ and the 
temperature inside the reactor. 
The net production rates are given as,
\begin{equation}
    \begin{bmatrix} 
    R_\Rm{Ee} \\
    R_\Rm{Ea} \\ 
    R_\Rm{A} \\ 
    R_\Rm{V} \\
    R_\Rm{W} \\ 
    R_\Rm{O} \\ 
    R_\Rm{C}  
    \end{bmatrix}
    =
    \begin{bmatrix} 
    -1 & - 1 \\
    0 & 0 \\
    -1 & 0 \\
    1 & 0 \\
    1 & 2 \\
    -(1/2) &  -3 \\
    0 & 2
    \end{bmatrix}
    \begin{bmatrix} 
    r_1 \\ 
    r_2 
    \end{bmatrix}
    \label{eq:netprodrate}
\end{equation}
We model the reactor as a single continuous stirred-tank reactor (CSTR) operating
isothermally with negligible pressure drop. The reactor is maintained at the same 
temperature as inlet stream 2 in \cref{fig:block_measurement}, whose temperature 
is a setpoint we specify.
We assume that the thermodynamic equation of state is valid even when the 
reactor is not at equilibrium \citep[pg. 133]{rawlings:ekerdt:2020}. Assuming an ideal 
gas mixture, the component balances for species $j$ in the reactor are:
\begin{align}
    \epsilon V_R \frac{dc_j}{dt} &= Q_f c_{jf} - Q c_j + (1 - \epsilon)V_R \rho_c R_j \notag\\
    Q &= Q_f+ \frac{\mathrm{R} T}{P} (1-\epsilon) V_R \rho_c \sum_{j} R_{j} \label{eq:mass_bal}
\end{align}
We make the following
assumptions to simplify the flowsheet: ethane, ethylene, oxygen and 
carbon dioxide are non-condensable and thus, we get perfect condensation of only
acetic acid, water and vinyl acetate in the flash drum. We also assume
that the separation in the distillation column is sharp. Thus, stream 
10 carries all the vinyl acetate and water produced in the reactor while stream 
8 contains only acetic acid. Hence, neither vinyl acetate nor water gets recycled
to the mixer. The \ce{CO2} absorber is assumed to remove only carbon
dioxide from stream 5. Finally, for unit operations other than the reactor, 
only material balances are solved; energy balances are not considered.

Let $ N_j^k $ be the molar flow of chemical species $j \in J 
\coloneqq \{ {\rm Ee, Ea, A, V, W, O, C} \}$ in stream $k \in K \coloneqq
\{1, 2, \ldots, 13 \}$. Hence, the molar flow coming out of the reactor can be 
written as $N_j^3 = Q c_j$. Since only the reactor contributes to dynamics in the flowsheet,
all other process units are assumed to operate at \textit{quasi} steady-state. 
Thus, the 
material balance for the mixer at any time can be 
concisely represented as:
\begin{align}
    N^2 &= N^1 + N^7 + N^{13} = N^1 + \diag(\zeta) \cdot N^3 \notag\\
    \zeta &= (\alpha, \alpha, \beta, 0, 0, \alpha, \alpha (1 - \mu))^\Rm{T} \notag\\
    Q_f &= \sum_j N_j^2 RT / P \label{eq:mixer_bal}
\end{align}
where $\alpha \coloneqq N_j^7 / N_j^6, \beta \coloneqq N_j^{13}/N_j^{8}$ are the
recycle fractions corresponding to streams 7 and 13 respectively,
while $\mu \coloneqq N_\Rm{C}^{11}/N_\Rm{C}^{5}$
is the fraction of molar flow of carbon dioxide removed by the \ce{CO2} absorber.
The concentrations at the reactor inlet can be calculated using $c_{jf} = N_j^2 / Q_f$. 
\Cref{eq:rxnrate,eq:netprodrate,eq:mass_bal,eq:mixer_bal} describe
the transient behavior of the flowsheet forming a differential algebraic equation
(DAE) system. We can simplify the DAE into an ODE by solving \cref{eq:mixer_bal} 
for $Q_f$ in terms of $c_j$ only. For the steady-state of the flowsheet, we need only 
to set the RHS terms 
of the ODE equations equal to zero. Note that the feeds $N_\Rm{W}^1, N_\Rm{V}^1, N_\Rm{C}^1$
are set to zero while the inert feed enters as $N_{\Rm{Ea}}^1 = \lambda N_{\Rm{Ee}}^1$ where $\lambda$ 
is a small constant number. Constructing the remaining streams in the flowsheet
is described in \Cref{app:process_units}.

\subsection{Economic optimization}
\label{sec:VAc_econ_opt}
A candidate objective function that takes into account the steady-state economics 
of the flowsheet is defined as follows,
\begin{align}
    \ell(x_s, u_s; p) \coloneqq{}& \small{p_1 N_V^{10} - p_2 \sum_j N_j^7 - p_3 \sum_j N_j^{13}} \notag \\
                      &- \small{\sum_j p_{4j} N_j^1 - p_5 \sum_j N_j^9 - p_6 \sum_j N_j^{12}}
\end{align}
which is used to formulate the following RTO problem:
\begin{align}
    \max_{x_s, u_s}\ell(x_s, u_s; p) & \notag \\
    \textnormal{s.t.} \quad f(x_s, u_s) ={}& 0 \notag \\
    g(x_s, u_s) ={}& N^{10}_{V, \Rm{nom}} - N_V^{10} \leq 0 \notag \\
    \underline{u} \leq u_s \leq \bar{u}{}& \label{eq:econ_opt}
\end{align}
where $x$ are the states which are the species concentrations $c$ in the reactor
and $c \coloneqq(c_{\Rm{Ee}}, c_{\Rm{Ea}}, c_\Rm{A}, c_\Rm{V}, c_\Rm{W}, c_\Rm{O}, c_\Rm{C})^\Rm{T}$.
The decision variables (or optimization degrees of freedom) $u$ are the feed rates,
recycle fractions and the reactor inlet temperature,
\begin{equation*}
u \coloneqq[N^1_{\Rm{Ee}}, N^1_{\Rm{O}}, N^1_{\Rm{A}}, \alpha, \beta, T]^\Rm{T}
\end{equation*}
The price vector
$p \coloneqq(p_1, p_2, p_3, p_{4j}^\Rm{T}, p_5, p_6)^\Rm{T}$ characterizes the 
profit function $\ell(\cdot)$. The RHS of the ODE formed 
by \cref{eq:rxnrate,eq:netprodrate,eq:mass_bal,eq:mixer_bal} is represented by $f(\cdot)$.
The constraint $g(\cdot)$ ensures 
that the nominal production rate of vinyl acetate is met. We set 
$N^{10}_\mathrm{V,nom} = 0$ to make the problem unconstrained.
We also fix the liquid recycle $\beta = 0.6$ to remove the flat direction
it forms with $N^1_\Rm{A}$. The vector $p$ and the bounds on 
the degrees of freedom $\underline{u}, \bar{u}$ are tabulated in \Cref{app:parameter_values}.
The subscript $s$ denotes the steady-state values of the states and the 
degrees of freedom.

The meaning of the terms in the objective function is 
as follows:
\begin{itemize}
    \item $p_1$ represents the \emph{revenue} from the vinyl acetate production $N_V^{10}$.
    While the remaining costs represent the overall \emph{operating expenditure} (opex).
    \item $p_2, p_3$ represent energy processing costs for gas and liquid recycle 
    streams $N^7, N^{13}$, while $p_5, p_6$ represent the cost to process the purge 
    streams $N^9, N^{12}$.
    \item $p_{4j}$ represent the material costs associated with each species $j$ in the 
    fresh feed $N^{1}$.
\end{itemize} 
If we supply excessively high feed rates, the residence time in the reactor
and consequently the yield of vinyl acetate will be low. On the other hand, very low
feed rates will lead to low production rates. A high gas recycle
can lead to a better yield but incurs a large energy cost.
Thus, an unconstrained optimum should exist in terms of the degrees of freedom
for the above objective function.
Solving \cref{eq:econ_opt}, we obtain
\begin{equation*}
u_s^\star = [15.54 \ \unitfrac{mol}{s}, 11.14 \ \unitfrac{mol}{s}, 17.15  \ \unitfrac{mol}{s}, 0.57, 0.6, 427.1 \ \Rm{K}]^\Rm{T}
\end{equation*}
with 
\begin{equation*}
\ell^\star = 292.87 \ \Rm{\$/s} \qquad N_V^{10}=13.13 \ \unitfrac{mol}{s}
\end{equation*} 
We analyze the reduced Hessian of the profit function with respect to the 
degrees of freedom. Because $u$ contains variables with different units, 
we rescale them using $\tilde{u}$ with 
$\tilde{u} = u / s_u$, where the scale factors $s_u$ are tabulated 
in \Cref{app:parameter_values}, so that the condition number reflects curvature
rather than the choice of units. The eigenvalues of the scaled reduced 
Hessian $\dfrac{\partial^2 \ell}{\partial \tilde{u}_s^2}$ are all negative, 
confirming a local maximum, and its condition number is 415, 
so the optimum is locally well-conditioned with no flat directions.
We performed a multi-start search from different initial guesses
for $u_s$ and $x_s$. It revealed no other local optima within
$\underline{u} \leq u_s \leq \bar{u}$. Together with the negative definite scaled
reduced Hessian, this establishes a strict, well-conditioned local maximum with
no competing optimum detected in the feasible region. At the computed optimum, 
profit is $7.4\%$ of revenue, a margin typical of commodity chemicals, so the 
optimum reflects a genuine economic trade-off rather than a tiny difference 
between two large numbers.

We retain the decision variable scaling for the models developed in the upcoming sections.
Problem \cref{eq:econ_opt} is solved using the IPOPT solver provided by CasADi 
\citep{wachter:biegler:2006, andersson:gillis:horn:rawlings:diehl:2019}.

\subsection{Structured models}

We consider \cref{eq:rxnrate,eq:netprodrate,eq:mass_bal,eq:mixer_bal} 
as the \emph{ground truth} for the flowsheet. In reality, however, some 
parts of the plant model are not known precisely. The reactor is where the chemical 
transformation takes place, and the kinetics are usually not known from first principles.
This is almost always the case in industrial practice \citep{bindlish:2025}. 

We assume that the separation factors for the simple separators are known. 
Thus, we choose to parameterize 
only the rate laws using feedforward neural networks as follows:
\begin{align}
    r_1 &= r_{\Rm{NN},1}(c_{\Rm{Ee}}, c_{\Rm{O}}, c_{\Rm{A}}, T; \theta_{r_{\Rm{NN},1}}) \notag\\
    r_2 &= r_{\Rm{NN},2}(c_{\Rm{Ee}}, c_{\Rm{O}}, T; \theta_{r_{\Rm{NN},2}}) 
    \label{eq:rate_nn}
\end{align}  
where $\theta_{r_{\Rm{NN},1}}, \theta_{r_{\Rm{NN},2}}$ represent the weights 
and biases
of the neural networks.
These rate laws replace \cref{eq:rxnrate} in the plant model and are trained 
as follows. 

If we represent the ODE formed by \cref{eq:mass_bal,eq:mixer_bal} 
by $\dot{x} = f(x, u)$, then we can write the following optimization problem 
to train
the neural-network parameters $\Theta_\Rm{NN} \coloneqq
(\theta_{r_{\Rm{NN},1}}, \theta_{r_{\Rm{NN},2}})$ as,
\begin{align}
	\Theta^\star_\Rm{NN} &= \argmin_{\Theta_\Rm{NN}}
		 \left( \frac{1}{N_{tr}N_t} \sum_{i=1}^{N_{tr}} \sum_{k=0}^{N_t}
          \norm{ \frac{y_i(k) - \yhat_i(k)}{\sigma_y} }^2 \right)^{1/2} \notag \\
		\textnormal{s.t.} \quad \dot{\hat{x}}_{i} &= f(\hat{x}_{i}, u_i; \Theta_\Rm{NN}) \notag \\ 
		\yhat_i &= h(\hat{x}_{i}, u_i) \notag \\
        \hat{x}_i(0) &= f_0(y_i(0)) 
        \label{eq:struc_train}
\end{align}	
where $x, u$ have been defined in \cref{sec:VAc_econ_opt} and $y = x$, thus $h(x, u) = \Rm{I}$.
$N_t$ represents the number of time-steps in a trajectory while $N_{tr}$ represents 
the number of trajectories in the training dataset. Training over a 
prediction horizon ($N_t$) reduces the compounding of one-step errors 
during open-loop rollout \citep{kumar:rawlings:2023a}.
The symbol $\sigma_y$ represents the
standard deviation of the measurements $y$ over the training dataset, while
the hat represents model predictions.
For full-state measurement, we simplify the 
training by initializing each trajectory at its first measurement rather than 
estimating the initial conditions, i.e., $f_0(y_i(0)) = y_i(0)$.

Since the model shown in \cref{eq:struc_train} enforces both mass conservation 
and the equation of state, we call it a \emph{structured} model. As written,
problem \cref{eq:struc_train} is the most realistic formulation in an industrial
setting, where only concentration measurements are available.
However, we consider one more way to train the networks in \cref{eq:rate_nn}.

Suppose that we have access to the \emph{rate} measurements $[r_1, r_2]^\Rm{T}$ over 
the same
training dataset used for \cref{eq:struc_train}, thus we can 
train the neural networks simply as,
\begin{align}
    \Theta^\star_\Rm{NN} &= \argmin_{\Theta_\Rm{NN}}
        \left( \frac{1}{N_{tr} N_t} \sum_{i=1}^{N_{tr}} \sum_{k=0}^{N_t}
          \norm{ \frac{r_i(k) - \hat{r}_i(k)}{\sigma_r} }^2 \right)^{1/2}
          \label{eq:struc_train_rates}
\end{align}
where $\sigma_r$ is the standard deviation of the rate measurements over the 
training dataset. 

We do not know the neural-network parameters that reproduce 
the plant kinetics \emph{exactly}. A rate measurement reflects the kinetic rate 
itself, rather than its integrated effect on concentration through the ODE 
in \cref{eq:struc_train}.
Hence \cref{eq:struc_train_rates} gives the closest estimate of the closure we 
can obtain from data. Training on rates therefore acts as both a reference 
and a diagnostic. It tests whether the network has the capacity to represent 
the kinetics and whether the resulting model recovers the plant's RTO optimum. 
If it does, we treat the rate-trained network as the \emph{best} attainable surrogate 
for the true closure and judge the concentration-trained models against it. 
If it does not, the shortfall lies with the dataset rather than the training, 
since no measurement is more informative about the kinetics than the rates 
themselves, and the dataset must be \emph{enlarged}.
Thus using \cref{eq:struc_train,eq:struc_train_rates} we consider three cases, 
defined together with all other cases studied in \cref{table:cases}:
\begin{enumerate}[label=(\alph*)]
    \item \textit{rmeas}: Problem \cref{eq:struc_train_rates} with access to rate 
    measurements $[r_1, r_2]^\Rm{T}$.
    \item \textit{cmeas\_rinit}: Problem \cref{eq:struc_train} \textit{initialized} with 
    the NN parameters learned from \textit{rmeas}.
    \item \textit{cmeas}: Problem \cref{eq:struc_train} with access to only 
    concentration measurements $c$.
\end{enumerate}

\subsection{Black-box models}

We also train a complete black-box model for the reactor as,
\begin{align}
	\theta^\star_{f_{\Rm{NN}}} &= \argmin_{\theta_{f_{\Rm{NN}}}}
		 \left( \frac{1}{N_{tr} N_t} \sum_{i=1}^{N_{tr}} \sum_{k=0}^{N_t}
          \norm{ \frac{y_i(k) - \yhat_i(k)}{\sigma_y} }^2 \right)^{1/2} \notag \\
		\textnormal{s.t.} \quad \dot{\hat{x}}_{i} &= f_{\Rm{NN}}(\hat{x}_{i}, u_i; \theta_{f_{\Rm{NN}}}) \notag \\ 
		\yhat_i &= h(\hat{x}_{i}, u_i) \notag \\
        \hat{x}_i(0) &= f_0(y_i(0))
        \label{eq:blackbox_nn}
\end{align}	
where the RHS $f$ of the ODE is approximated using a neural network. 
As in the previous section,
we use $y=x$ and set $f_0(y_i(0)) = y_i(0)$.
By construction, \cref{eq:blackbox_nn} preserves neither individual species mass 
conservation nor the equation of state inside the reactor; it learns them from 
data instead.
Hence, we call it a \textit{black-box} model. 
To use it for RTO, we need to calculate the flowsheet streams.
For that, we need to know the volumetric flow $Q$ coming out of the reactor.
From \cref{eq:mixer_bal}, by applying an overall mass-balance across the flowsheet 
at steady state we get,
\begin{align}
    Q = \frac{\sum_{j} M_j N^1_j}{\sum_{j} M_j (1 - \zeta_j) c_j} \label{eq:Q_nn}
\end{align}
where $M_j$ is the molecular weight of the species $j \in J$.
Thus, using \cref{eq:blackbox_nn,eq:Q_nn} and $N^3_j = Qc_j$ we can construct 
the flowsheet and use
the black-box model for economic optimization.

\section{Training and implementation details}
\label{sec:training}

\subsection{Training dataset}
As shown in \cref{fig:block_measurement}, we record the species concentrations 
at the reactor outlet while perturbing the decision variables with pseudo-random
binary signals under a zero-order hold. The decision variables
are the feed rates $N^1_\Rm{Ee}, N^1_\Rm{O}, N^1_\Rm{A}$, the
recycle fractions $\alpha, \beta$ and the reactor inlet temperature $T$. 
We sample 200 random inputs and hold out 30\% as a validation set. 
The inputs $u$, except temperature, are varied by $\pm 30\%$ around their optimal
values $u_s^\star$, while temperature is varied by $\pm 10\%$. These limits
coincide exactly with the box constraints of the RTO problem \cref{eq:econ_opt}, i.e., 
$\underline{u} = 0.7\,u_s^\star$ and $\bar{u} = 1.3\,u_s^\star$ for
all the feed rates and recycle fraction $\alpha$, and $\underline{u} = 0.9\,T^\star$,
$\bar{u} = 1.1\,T^\star$ for the temperature, tabulated in \Cref{app:parameter_values}.
Although $\beta$ is perturbed $\pm 30\%$ to excite the data, it is held fixed at $0.6$ in 
the RTO problem \cref{eq:econ_opt} (i.e., $\underline{u} = \bar{u}$ for $\beta$).
The training dataset therefore spans exactly the feasible box over which we optimize.
These perturbations are more aggressive than industrial model identification
typically permits, so the training data here is richer than a practitioner could likely obtain.

\begin{figure*}[!tp]
    \centering
    \includegraphics[width=\textwidth, page=1]{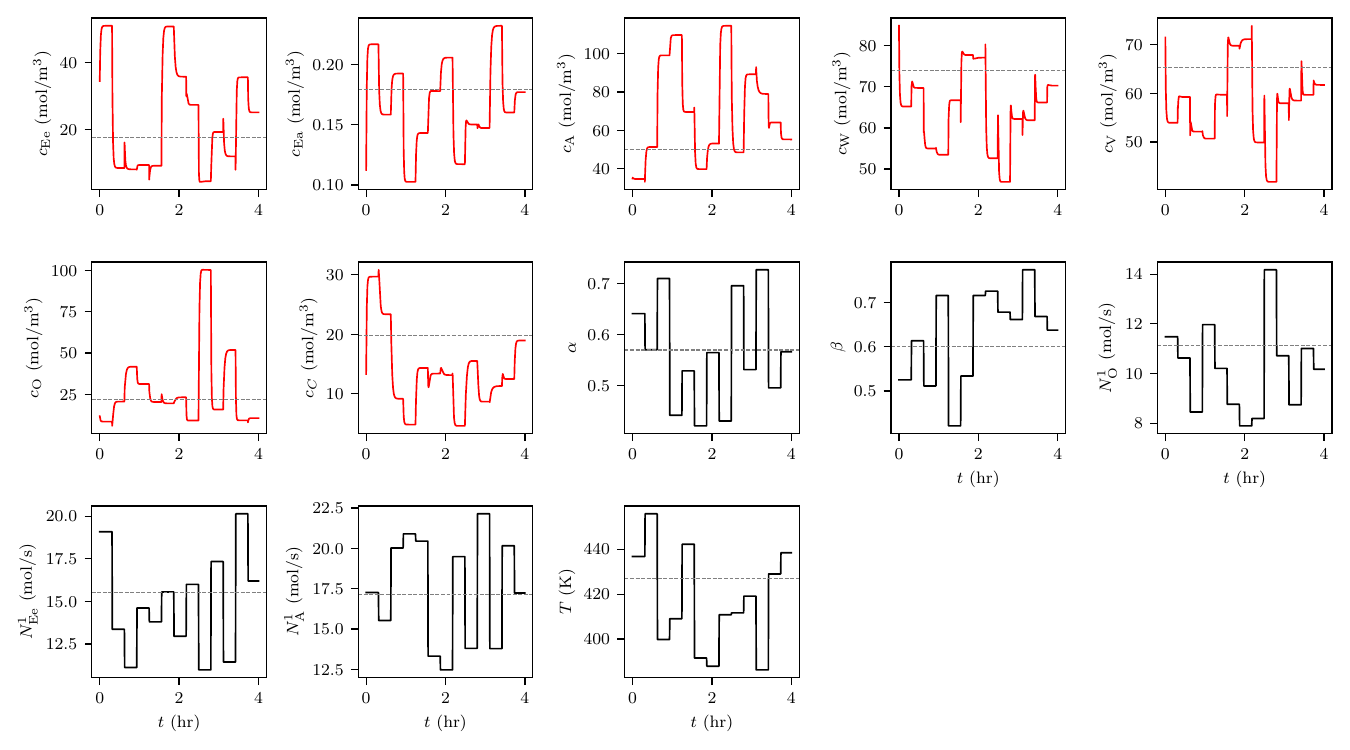}
    \caption{A snapshot of training data from the plant model with no 
            measurement noise. The measurements are 
            shown in red and the decision variables in black.
            The dotted grey lines represent $x_s^\star$ and $u_s^\star$ from \cref{sec:VAc_econ_opt}.}
    \label{fig:training_data}
\end{figure*}

A subset of the noise-free training dataset is shown in 
\cref{fig:training_data}, and it brackets the measurements at the steady-state optimum $x_s, u_s$. 
Thus, we test the models purely on \textit{interpolation} and not extrapolation.
We ensure that the training data has enough steady-state information by holding 
the decision variables constant for a fixed period before injecting the next random step,
so that the model learns the steady-state behavior of the flowsheet
\citep{kumar:rawlings:2023a}.

\subsection{Software details}
Both the \emph{structured} and \emph{black-box} models are built and trained using the \texttt{JAX} 
open-source Python library developed by Google DeepMind \citep{bradbury:frostig:hawkins:johnson:et_al:2018}. 
We solve the ODEs using \texttt{Diffrax} \citep{kidger:2021} with an adaptive 
step-size solver of the ESDIRK5(4) type. The neural networks are constructed and initialized
using \texttt{Flax} \citep{flax:2020}. 

The hyperparameters in this paper are the network width and depth, the learning rate, 
the batch size, and the epoch budget. We fixed the width and depth because 
the training error was already low at this architecture, so additional capacity 
was unnecessary. We also fixed the batch size and tuned 
only the learning rate and the epoch budget. We first swept the learning rate 
and kept the value with the lowest validation loss, then increased 
the epoch budget until the validation loss plateaued.
The tuning procedure for each model we construct is detailed in 
\crefrange{fig:lr_cmeas}{fig:lr_bbox_noise} of the supplementary material.
We run ADAM \citep{kingma:ba:2014} for the first half of the epoch budget
and then switch to L-BFGS \citep{liu:nocedal:1989} for the second
half for faster convergence. A comparison of using only ADAM versus
ADAM followed by L-BFGS is also shown in the supplementary material.
Both the optimizers 
are obtained from \texttt{Optax} provided by JAX.

\subsection{Scaling and initialization}

We use hyperbolic tangent as the activation function for the neural networks. 
Since both \textit{structured} and \textit{black-box} models are used inside a 
nonlinear program (NLP) \cref{eq:econ_opt},
\textit{tanh} ensures differentiability
of the model outputs with respect to the inputs and parameters. 

For \textit{tanh}, both the inputs and outputs of the neural networks should be 
scaled so that the
network operates on $\mathcal{O}(1)$ quantities. The inputs are normalized by
their mean and standard deviation over the entire training dataset. For the
\textit{structured} model, we deliberately separate the \emph{shape} of the rate function from
its \emph{scale}: the network learns the standardized shape, while a trainable
location $\hat{\mu}_{r}$ and scale $\hat{\sigma}_{r}$ restore the physical magnitude. For example,
$r_1$ is scaled as
\begin{align*}
    r_1 = \hat{\mu}_{r_1} + \hat{\sigma}_{r_1} \times r_{\Rm{NN},1}(\cdot),
\end{align*}
so that $r_{\Rm{NN},1}(\cdot)$ need only output an $\mathcal{O}(1)$ standardized
rate, while $\hat{\mu}_{r_1}$ and $\hat{\sigma}_{r_1}$ carry its mean and spread. 
This scaling keeps the
raw network output well-scaled and the back-propagated gradients
well-conditioned; without it, the network would have to emit values spanning
the raw rate magnitudes directly, which readily causes \textit{vanishing} or
\textit{exploding} gradients, especially for stiff systems \citep{fronk:petzold:2025}. 
To further stabilize training, we apply gradient
clipping to the network parameters $\Theta_\Rm{NN}$, but leave the scale factors
$\hat{\mu}_{r_i}, \hat{\sigma}_{r_i}$ unclipped. The weights are thereby protected from large,
destabilizing updates, while $\hat{\mu}_{r_i}$ and $\hat{\sigma}_{r_i}$ remain free to take the larger
steps needed to reach the correct rate magnitude. We apply clipping only during 
the stochastic ADAM phase. The deterministic L-BFGS phase needs none, since its 
strong Wolfe line search already bounds each step size.

The factors $\hat{\mu}_{r_i}, \hat{\sigma}_{r_i}$ must therefore be initialized with care.
For problem \cref{eq:struc_train}, we
initialize at the mean and standard deviation of $[r_1, r_2]^\Rm{T}$ 
computed from
the plant model over the training dataset, so that the network begins by
learning only the standardized shape. In practice, the plant rates are unavailable.
One can instead use \cref{eq:blackbox_nn} to get the RHS and back-calculate
approximate rates using \cref{eq:mass_bal} to set these initial estimates as shown
by \cite{peng:liq:boy:2025}, though
that is not the focus of this work.
For problem \cref{eq:struc_train_rates}, by contrast, we do not need these scale factors, since
they are already available from the assumed rate measurements.

For the \textit{black-box} model, we use a similar scaling strategy. The inputs 
to the neural network $f_\Rm{NN}$ are normalized by their mean and standard deviation over 
the training dataset. The outputs are scaled by trainable location and scale 
factors $\hat{\mu}_{f_\Rm{NN}}, \hat{\sigma}_{f_\Rm{NN}}$ that restore the physical 
magnitude while the network learns the standardized shape of the RHS of the ODE. 
As done for the structured model, we initialize $\hat{\mu}_{f_\Rm{NN}}$ and $\hat{\sigma}_{f_\Rm{NN}}$ 
at the mean and standard deviation of the true RHS $\dot{x}$ computed from
the plant model over the training dataset.

\begin{table*}[!tp]
    \centering
    \caption{Summary of the six data-driven cases. $n_l$ is the
    sensor-noise level in \cref{eq:noise}. A random NN initialization uses the LeCun 
    normal initialization popular for \textit{tanh}-based neural networks 
    \citep[pg. 20]{lecun:leon:orr:muller:2002}.}
    \label{table:cases}
    \begin{tabular}{llccc}
        \toprule
        Model & Case & Training problem and measurements & NN initialization & Noise $n_l$ \\
        \midrule
        \multirow{4}{*}{Structured}
          & \textit{rmeas}       & \cref{eq:struc_train_rates}, rates    & random    & $0$ \\
          & \textit{cmeas\_rinit} & \cref{eq:struc_train}, concentrations & \textit{rmeas} solution & $0$ \\
          & \textit{cmeas}       & \cref{eq:struc_train}, concentrations & random    & $0$ \\
          & \textit{cmeas\_noise}& \cref{eq:struc_train}, concentrations & random    & $10^{-2}$ \\
        \midrule
        \multirow{2}{*}{Black-box}
          & \textit{bbox}        & \cref{eq:blackbox_nn}, concentrations & random & $0$ \\
          & \textit{bbox\_noise} & \cref{eq:blackbox_nn}, concentrations & random & $10^{-2}$ \\
        \bottomrule
    \end{tabular}
\end{table*}

\section{Results and discussion}
\label{sec:results}

We demonstrated in \cref{sec:VAc_econ_opt} that the flowsheet defined 
by \cref{eq:rxnrate,eq:netprodrate,eq:mass_bal,eq:mixer_bal} 
has a well-conditioned optimum. 
As a control experiment, we first verify that the 
plant rate structure is identifiable from the data the plant generated, and that the 
identified model \emph{recovers} the plant optimum. 
This rules out uninformative data or an ill-posed problem
as a trivial explanation 
for any failure of the structured or black-box models. Whether the optimum 
remains recoverable without the \emph{correct structure} is precisely the 
question these experiments address.

\subsection{Control experiment using parametric model}

From \cref{eq:rxnrate}, we note that the plant rate laws are power law models.
We rescale and parameterize them as follows:
\begin{align}
    r_1 ={}& \tilde{k}_1
            \times e^{-\tilde{E}_1/R (1/T - 1/T_m)} \notag \\
          &\times (c_\Rm{Ee}/c_\Rm{Ee, m})^{a_1} (c_\Rm{O}/c_\Rm{O, m})^{b_1}
            (c_\Rm{A}/c_\Rm{A, m})^{c_1} \notag \\
    r_2 ={}& \tilde{k}_2
            \times e^{-\tilde{E}_2/R (1/T - 1/T_m)} \notag \\
          &\times (c_\Rm{Ee}/c_\Rm{Ee, m})^{a_2} (c_\Rm{O}/c_\Rm{O, m})^{b_2}
    \label{eq:rxnrate_param}
\end{align}
where $T_m, c_{j, m}$ are the mean temperature and 
concentrations over the subset used to fit the parametric model \cref{eq:rxnrate_param}. 
This subset is roughly $0.25\%$ of the entire training-and-validation dataset 
used for the neural networks, shown in \cref{fig:training_data}. From it we estimate the parameters 
$[\log{\tilde{k}_1}, \log{\tilde{E}_1}, a_1, b_1, c_1, \log{\tilde{k}_2}, \log{\tilde{E}_2}, a_2, b_2]$
along with their confidence intervals. 
To avoid biasing the estimator, we initialize far from the true plant parameters. 
We also corrupt each
measurement with multiplicative Gaussian sensor noise,
\begin{equation}
    \tilde{c} = c\,(1 + \eta\, n_l) = c + \eta\, n_l\, c,
    \label{eq:noise}
\end{equation}
where $\eta \sim \mathcal{N}(0,1)$ is drawn independently for each measurement
and $n_l \in \{0,\, 10^{-3},\, 10^{-2}\}$ is the relative error level
(0, 0.1\%, 1\%). The parameter estimation problem is solved using \texttt{Paresto}
\citep{dake:ilagan:banerjee:scott:rawlings:2024} which provides joint confidence 
intervals for the parameters. \texttt{Paresto} solves a similar problem as \cref{eq:struc_train}
 using IPOPT,
with $r_1, r_2$ replaced by \cref{eq:rxnrate_param} and $\Theta_\Rm{NN}$ by 
parameters in \cref{eq:rxnrate_param}. 
The results are shown in \cref{table:control_recovery}.
We see that the plant rate structure is identifiable from the training data, as 
indicated by the tight confidence intervals for the parameters even in the presence of noise.

Each of the fits reports a single optimum even on multi-start RTO. 
We measure this recovery with the following metric, henceforth
called the \emph{profit loss},
\begin{align}
    \Delta \ell = \frac{\ell_{\hat{u}_s} - \ell^\star}{\ell^\star} \times 100\%
\end{align}
where $\ell^\star$ is the optimal plant profit and
$\ell_{\hat{u}_s}$ is the profit obtained by evaluating the \emph{plant model}
at the estimated operating point $\hat{u}_s$ from an identified model.
Note that $\Delta \ell \leq 0\%$ since $\ell^\star$ is the global optimum. We see
from \cref{table:control_recovery} that the profit loss is negligible at all 
noise levels, which indicates that the
identified parametric model is able to recover the plant optimum. 
\begin{table*}[!tp]
  \centering
  \caption{Parameter estimation results and profit loss for the control 
  experiment with parametric models for different levels of measurement noise.}
  \begin{tabular}{ c r@{\,$\pm$\,}l r@{\,$\pm$\,}l r@{\,$\pm$\,}l }
    \toprule
    \thead{Parameter} &
    \multicolumn{2}{c}{\thead{$n_l = 0$}} &
    \multicolumn{2}{c}{\thead{$n_l = 10^{-3}$}} &
    \multicolumn{2}{c}{\thead{$n_l = 10^{-2}$}} \\
    \midrule
    $\log{\tilde{k}_1}$ & $-4.836$ & $2.7\times10^{-11}$ & $-4.840$ & $4.4\times10^{-3}$ & $-4.876$ & $3.9\times10^{-2}$ \\
    $\log{\tilde{E}_1}$ & $10.61$  & $4.3\times10^{-11}$ & $10.60$  & $7.1\times10^{-3}$ & $10.55$  & $6.7\times10^{-2}$ \\
    $a_1$               & $1.100$  & $5.2\times10^{-11}$ & $1.092$  & $8.6\times10^{-3}$ & $1.021$  & $7.7\times10^{-2}$ \\
    $b_1$               & $1.100$  & $5.5\times10^{-11}$ & $1.091$  & $9.0\times10^{-3}$ & $1.014$  & $8.1\times10^{-2}$ \\
    $c_1$               & $1.000$  & $4.4\times10^{-11}$ & $0.993$  & $7.2\times10^{-3}$ & $0.934$  & $6.5\times10^{-2}$ \\
    $\log{\tilde{k}_2}$ & $-7.992$ & $1.1\times10^{-11}$ & $-7.993$ & $1.8\times10^{-3}$ & $-7.994$ & $1.8\times10^{-2}$ \\
    $\log{\tilde{E}_2}$ & $10.87$  & $1.8\times10^{-11}$ & $10.87$  & $3.0\times10^{-3}$ & $10.87$  & $3.0\times10^{-2}$ \\
    $a_2$               & $1.000$  & $2.4\times10^{-11}$ & $0.9996$ & $4.0\times10^{-3}$ & $0.996$  & $4.0\times10^{-2}$ \\
    $b_2$               & $1.000$  & $4.0\times10^{-11}$ & $1.000$  & $6.7\times10^{-3}$ & $0.9997$ & $6.7\times10^{-2}$ \\
    \midrule
    $\Delta \ell$  & \multicolumn{2}{c}{$<-0.001\%$} & \multicolumn{2}{c}{$-0.019\%$} & \multicolumn{2}{c}{$-1.28\%$} \\
    \bottomrule
    \end{tabular}
    \label{table:control_recovery}
\end{table*}
These results can be used as a benchmark to compare
the performance of the structured and black-box models.

\subsection{Training and validation of the data-driven models}

\subsubsection*{(a) Structured models}
Because neural networks are over-parameterized and their training loss is
non-convex, each random initialization of the weights and biases can settle in
a different basin and recover a distinct rate function. For each of the three
cases \textit{rmeas}, \textit{cmeas\_rinit} and \textit{cmeas}, we therefore
train an ensemble of 20 networks, each from an independent random
initialization. Note that we randomly initialize only the NN weights and biases 
and \emph{not} the scaling factors. We train one additional
case called \textit{cmeas\_noise}, which is similar to \textit{cmeas} but uses 
noisy measurements \cref{eq:noise} with $n_l = 10^{-2}$ (i.e. $1\%$ relative 
sensor noise).

For the stochastic ADAM phase, the $N_{tr}$ trajectories are partitioned into $N_B$ \emph{batches}
of size $N_{tr}/N_B$, and the optimizer takes one gradient step per batch
(without resampling) usually to escape poor local minima/saddle points. One pass over 
all the batches is an
\emph{epoch}. We set $N_B = 5$ and run for 250 epochs.
We freeze the batch ordering across all 20 initializations to
remove shuffling as a confounder. Then, we switch to L-BFGS for 250 more epochs.
Note no batch partitioning is needed for L-BFGS, since it is a deterministic optimizer.

The architecture of the neural networks is also kept fixed across all cases.
Unlike \cref{eq:struc_train,eq:struc_train_rates}, which scale residuals by
$\sigma_y$, we report a mean-normalized RMSE on the measurement and validation
sets to read error as a fraction of the mean concentration,
\begin{equation}
    \Rm{RMSE}_{s} =
    \left( \frac{1}{N_s N_t}
        \sum_{i=1}^{N_s} \sum_{k=0}^{N_t}
        \norm{ \frac{y_{i}(k) - \hat{y}_{i}(k)}{\mu_{y}} }^2 \right)^{1/2}
    \label{eq:rmse_report}
\end{equation}
for $s \in \{\Rm{train}, \Rm{val}\}$
where $\mu_{y}$ is the mean output value, and
$N_s$ the number of trajectories in set $s$; e.g.\ $0.02$ means a $2\%$ error
relative to the mean concentrations.

\subsubsection*{(b) Black-box model}

We train the black-box model with the same strategy as the structured models, taking care to remove any
confounding factors across the 20 initializations. We label \cref{eq:blackbox_nn}
as \textit{bbox} and add a noisy-measurement case \textit{bbox\_noise}, analogous to
\textit{cmeas\_noise}. A summary of all six cases considered in the paper
is provided in \cref{table:cases}.
We set $N_B = 5$ and run 
for 250 epochs of ADAM followed by 250 epochs of L-BFGS.
We use the same mean-normalized RMSE of \cref{eq:rmse_report} to report 
the training and validation errors.

\subsubsection*{(c) Validation}
The training loss, validation loss and training time for all 
the identified models are reported in \cref{table:model_results}.

\begin{figure*}[!tp]
    \centering
    \includegraphics[width=\textwidth, page=15]{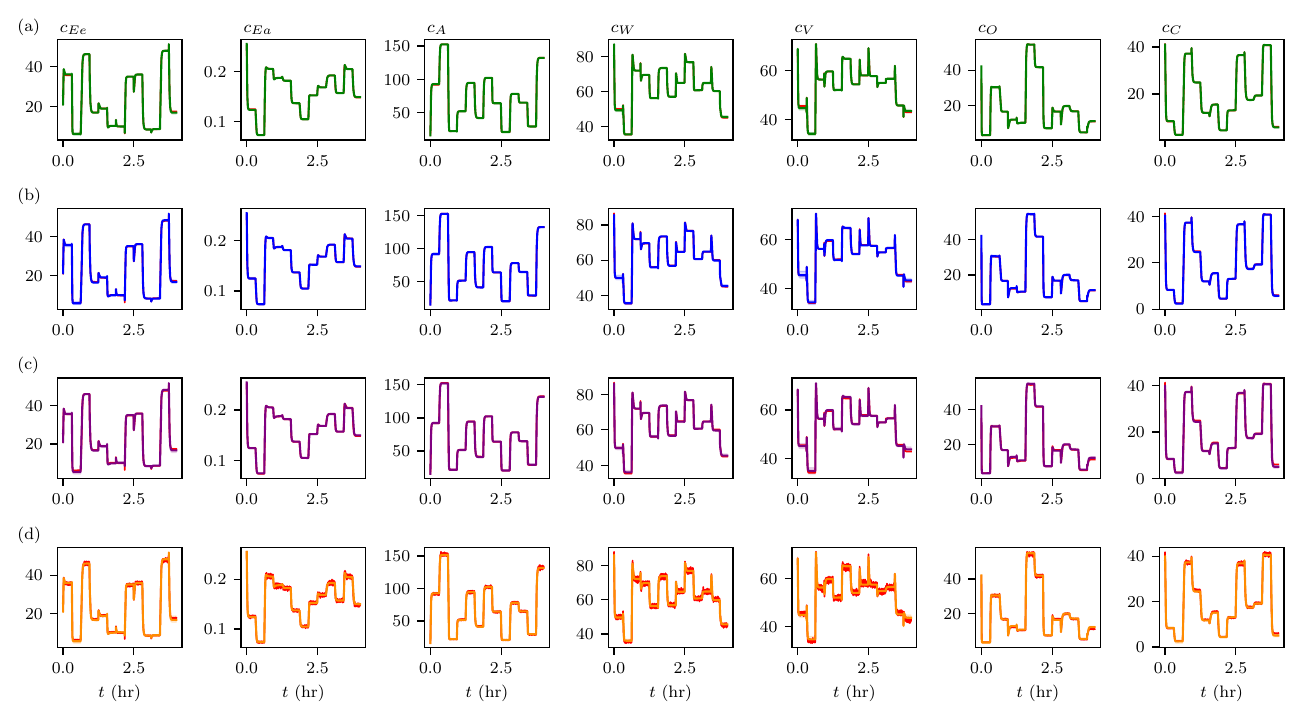}
    \caption{Concentration predictions on the validation set for (a) \textit{rmeas}, (b) \textit{cmeas\_rinit}, (c) \textit{cmeas} and (d) \textit{cmeas\_noise}.
    Measurements are shown in red. The shaded bands are the ensemble spread, taken as the 5--95\% quantiles of
    the 20 random initializations.}
    \label{fig:validation_plots_struc}
\end{figure*}

\begin{figure*}[!tp]
    \centering
    \includegraphics[width=0.95\textwidth, page=14]{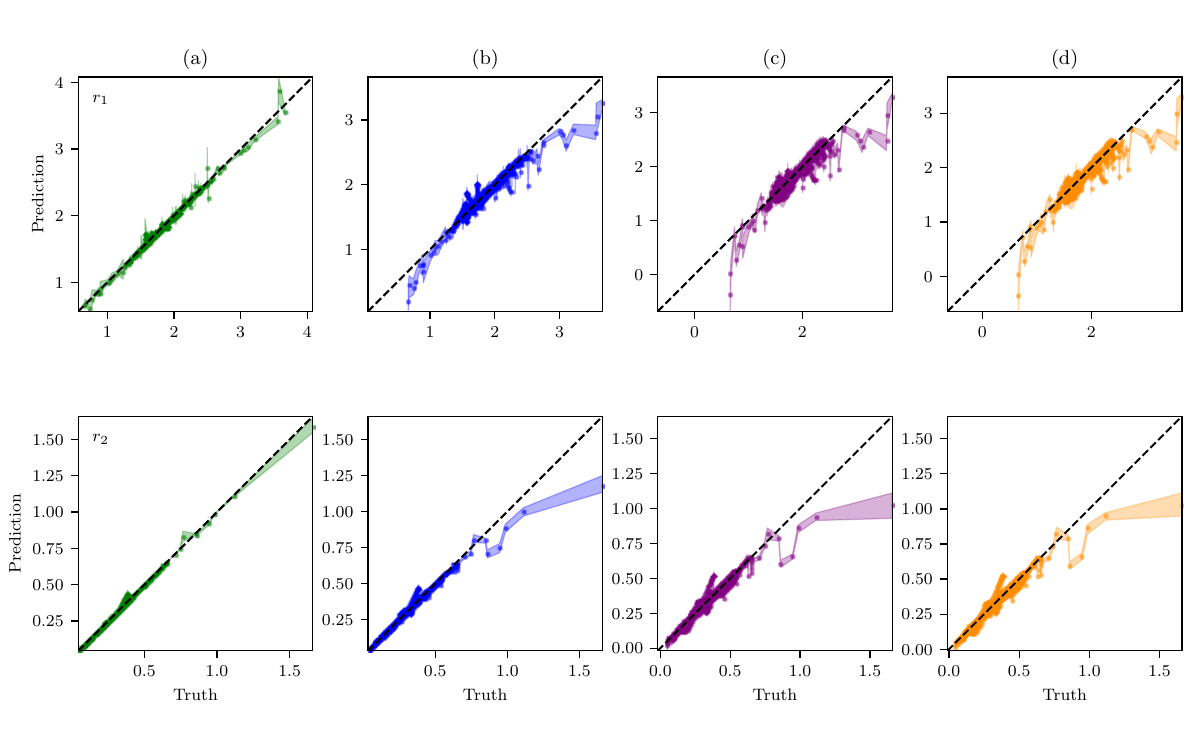}
    \caption{Parity plots of the reaction rates $[r_1, r_2]^\Rm{T}$ on the validation set for (a) \textit{rmeas}, (b) \textit{cmeas\_rinit}, (c) \textit{cmeas} and (d) \textit{cmeas\_noise}.
    Top row shows $r_1$ and bottom row $r_2$. The dotted line is ideal parity.
    The bands are the ensemble spread, taken as the 5--95\% quantiles across the 20 random initializations.}
    \label{fig:parity_plots}
\end{figure*}

For the structured models we first look at the quantity a practitioner can
actually observe, i.e., the concentration predictions on the validation set.
\Cref{fig:validation_plots_struc} overlays these predictions on the validation
measurements for all four cases. The fits are excellent, and the ensemble bands
are so tight that they are barely visible. Every random initialization produces
essentially the same predictions, so the 5--95\% quantile range collapses onto
the mean.
In our simulation setting, we can also inspect a quantity the practitioner never
sees, the plant reaction rates. The parity plots of the model rates against 
the truth in \cref{fig:parity_plots} tell a very different story. The rate
predictions degrade steadily from (a) \textit{rmeas} to (d) \textit{cmeas\_noise},
even though the concentration predictions above were indistinguishable.

\begin{figure*}[!tp]
    \centering
    \includegraphics[width=\textwidth, page=6]{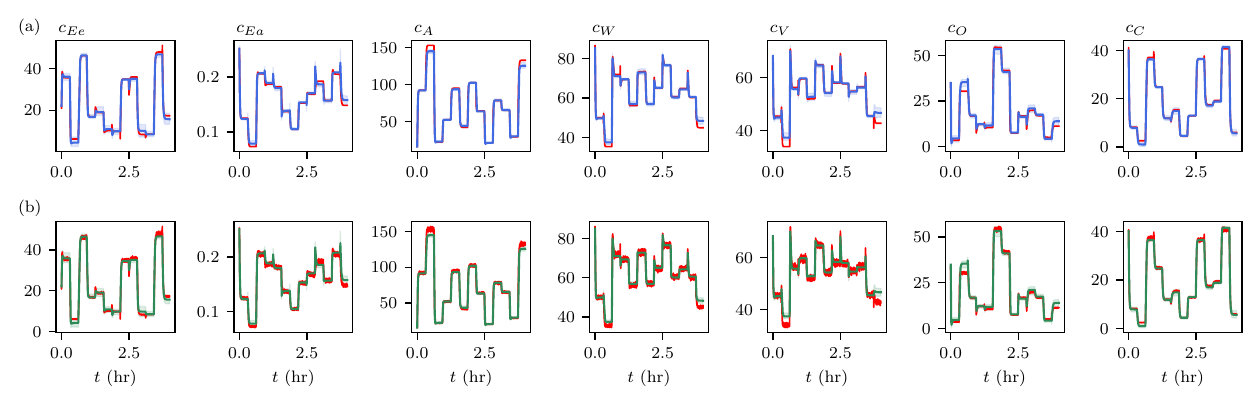}
    \caption{Concentration predictions on the validation set for the black-box model, (a) \textit{bbox} and
    (b) \textit{bbox\_noise}. Measurements are shown in red. The
    bands are the ensemble spread, taken as the 5--95\% quantiles of the 20 random
    initializations.}
    \label{fig:validation_plots_bbox}
\end{figure*}

Despite discarding all physical
structure, the black-box models attain a satisfactory fit when overlaid on the
validation measurements, as shown in \cref{fig:validation_plots_bbox}, 
though less accurate than that of the structured models. 
Here too the ensemble bands stay tight. The black-box models thus pass every check 
a practitioner would run.

\begin{table*}[!tp]
    \centering
    \caption{Structured- and black-box-model results across cases. Values are
    reported as mean~$\pm$~$1.96\,\sigma$ over 20 random initializations
    (95\% interval). \textsc{rmse}$_\Rm{meas}$ and \textsc{rmse}$_\Rm{val}$ are
    the mean-normalized errors of \cref{eq:rmse_report} on the measurement and
    validation sets; training time is wall-clock.}
    \label{table:model_results}
    \begin{tabularx}{\textwidth}{llcYYY}
        \toprule
        Model & Case & NN architecture & \textsc{rmse}$_\Rm{meas}$ & \textsc{rmse}$_\Rm{val}$ & Train time (hr) \\
        \midrule
        \multirow{4}{*}{Structured}
          & \textit{rmeas}       & \multirow{4}{*}{\makecell{$r_{\Rm{NN}, r_1} = [4, 64, 1]$ \\ $r_{\Rm{NN}, r_2} = [3, 64, 1]$}} & $0.00067 \pm 0.00012$ & $0.0069 \pm 0.0021$ & $0.22 \pm 0.08$ \\
          & \textit{cmeas\_rinit} &  & $0.00435 \pm 0.00060$ & $0.0146 \pm 0.0024$ & $11.97 \pm 0.94$ \\
          & \textit{cmeas}       &  & $0.00966 \pm 0.00131$ & $0.0219 \pm 0.0035$ & $12.27 \pm 1.17$ \\
          & \textit{cmeas\_noise}&  & $0.01472 \pm 0.00064$ & $0.0217 \pm 0.0030$ & $12.45 \pm 0.99$ \\
        \midrule
        \multirow{2}{*}{Black-box}
          & \textit{bbox}        & \multirow{2}{*}{$f_{\Rm{NN}} = [16, 128, 7]$} & $0.03224 \pm 0.00187$ & $0.0484 \pm 0.0051$ & $12.01 \pm 1.06$ \\
          & \textit{bbox\_noise} &  & $0.03410 \pm 0.00177$ & $0.0481 \pm 0.0057$ & $12.04 \pm 0.90$ \\
        \bottomrule
    \end{tabularx}
\end{table*}
Herein lies the trap. Every case we considered so far reports a low training and validation RMSE
and a tight ensemble spread across initializations, so nothing the practitioner
can compute from the data flags a problem. Yet, as we show next, five of these
six ``validated'' models are useless for RTO.

\subsection{Optimization landscape}
\label{sec:landscape}

Instead of solving the RTO problem \cref{eq:econ_opt} over all five degrees of freedom
(i.e. $[N_\Rm{Ee}^1, N_\Rm{O}^1, N_\Rm{A}^1,\alpha, T]^\Rm{T}$), 
we first solve a simpler problem over only two degrees of freedom. 
To see how each model reshapes the optimization landscape, we
vary only the reactor inlet temperature $T$ and the gas recycle fraction
$\alpha$ over their box constraints $[\underline{u}, \bar{u}]$ from
\cref{eq:econ_opt}, keeping the other three degrees of freedom fixed at the plant optimum.
The plant profit $\ell(T, \alpha)$ is then a 2D surface we can draw.

\begin{figure}[!htbp]
    \centering
    \includegraphics[width=\columnwidth, page=1]{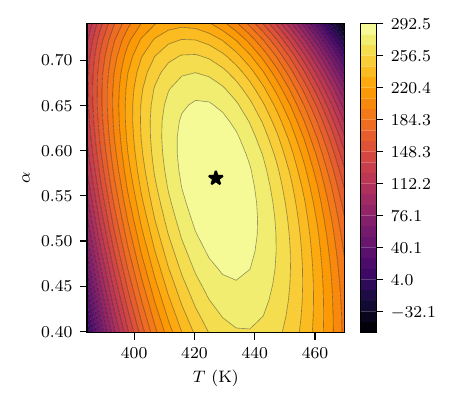}
    \caption{Plant profit landscape $\ell(T, \alpha)$ with the remaining degrees of freedom
    fixed at the plant optimum. The star marks the plant optimum
    $(T^\star, \alpha^\star)$. The surface is unimodal and well-conditioned.}
    \label{fig:landscape_plant}
\end{figure}

\begin{figure*}[!tp]
    \centering
    \includegraphics[width=\textwidth, page=22]{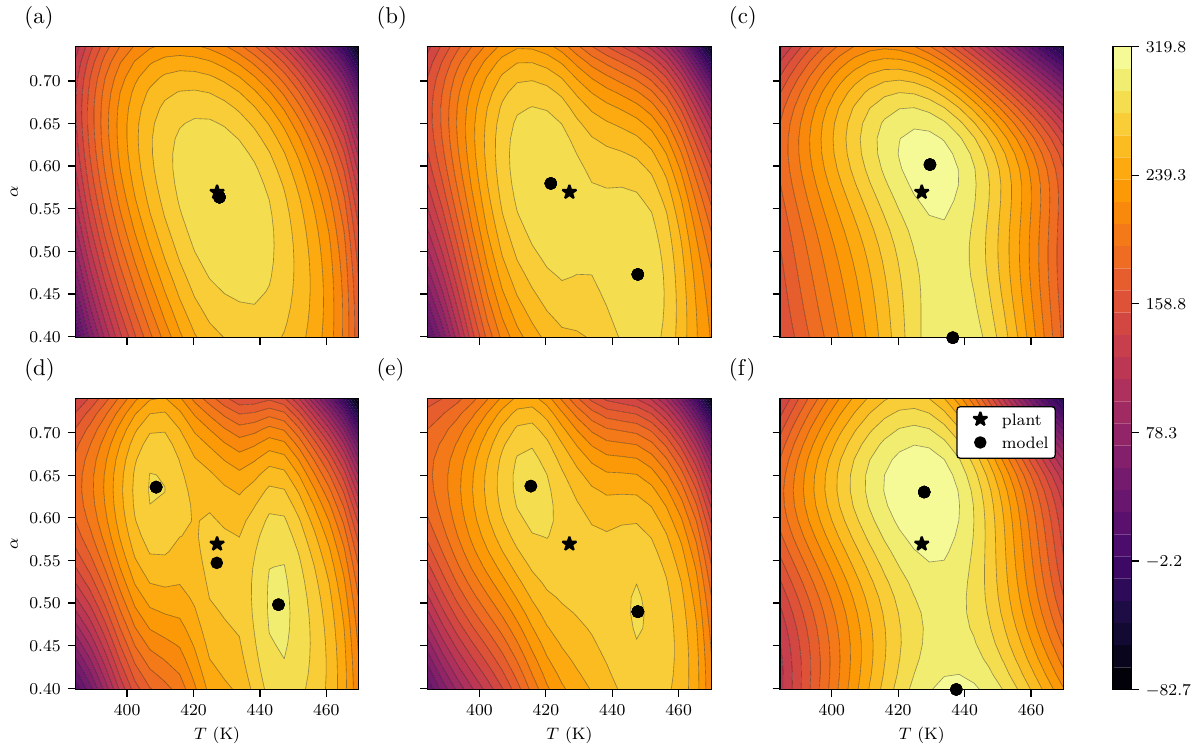}
    \caption{Model-predicted profit landscape $\hat{\ell}(T, \alpha)$ that each
    trained model presents to the RTO optimizer. Panels, in order of degrading structure and data,
    are (a)~\textit{rmeas}, (b)~\textit{cmeas\_rinit}, (c)~\textit{cmeas},
    (d)~\textit{cmeas\_noise}, (e)~\textit{bbox}, and (f)~\textit{bbox\_noise}.
    The star marks the plant optimum as a fixed reference and the circles each
    model's own optimum. Each panel shows one of the worst-performing members of the 20
    initializations.}
    \label{fig:landscape_models}
\end{figure*}

For model-based RTO, each trained
model operates on its own predicted profit $\hat{\ell}(T, \alpha)$. It is distinct 
from $\ell_{\hat{u}_s}$, the \emph{plant} profit evaluated at the
operating point the model recommends. This second quantity is what enters the 
profit-loss metric $\Delta \ell$ used later.
\Cref{fig:landscape_plant} shows the plant landscape $\ell$,
and \cref{fig:landscape_models} the model landscape $\hat{\ell}$ that each trained
model presents to the RTO optimizer, for one of the worst-performing members
of the ensemble of 20, on the same axes and profit scale.

The plant landscape in \cref{fig:landscape_plant} has a single peak at
$(T^\star, \alpha^\star)$, consistent with the well-conditioned optimum
discussed in \cref{sec:VAc_econ_opt}. 

\Cref{fig:landscape_models} tells a different story. Trained on rate measurements,
\textit{rmeas} reproduces both the plant landscape and its peak. 
The \textit{rmeas} case also helps us diagnose that the training dataset is informative enough to identify
the plant rate structure. However, \textit{cmeas\_rinit} (\cref{table:cases}) is the telling case.
Initialized at the same NN weights as \textit{rmeas} that recovers the plant
peak, the stochastic training optimizer deforms the surface so
that \textit{cmeas\_rinit} no longer agrees with the landscape it started from.
The same drift is visible in the loss curves in \Cref{fig:lr_cmeas_rinit} of the supplementary
material, where ADAM moves away from the initialization and L-BFGS recovers only
part of it. We trained \textit{cmeas\_rinit} at the same learning rate (i.e., 0.05) tuned for
\textit{cmeas} and \textit{cmeas\_noise}. A smaller learning rate does damp the
drift, only because it preserves an optimum that
was supplied, not one that training located. An optimizer should not move away
from an optimum simply because of a user-defined hyperparameter.
\textit{cmeas}, \textit{cmeas\_noise}, \textit{bbox} and \textit{bbox\_noise}
distort the surface further with multiple local maxima. \textit{cmeas} even admits an
optimum in a near-saddle region, where the curvature is indefinite and
gradient-correction schemes such as modifier adaptation would fail to converge.

\subsection{Economic performance under RTO}

\begin{figure*}[!tp]
    \centering
    \includegraphics[width=\textwidth, page=7]{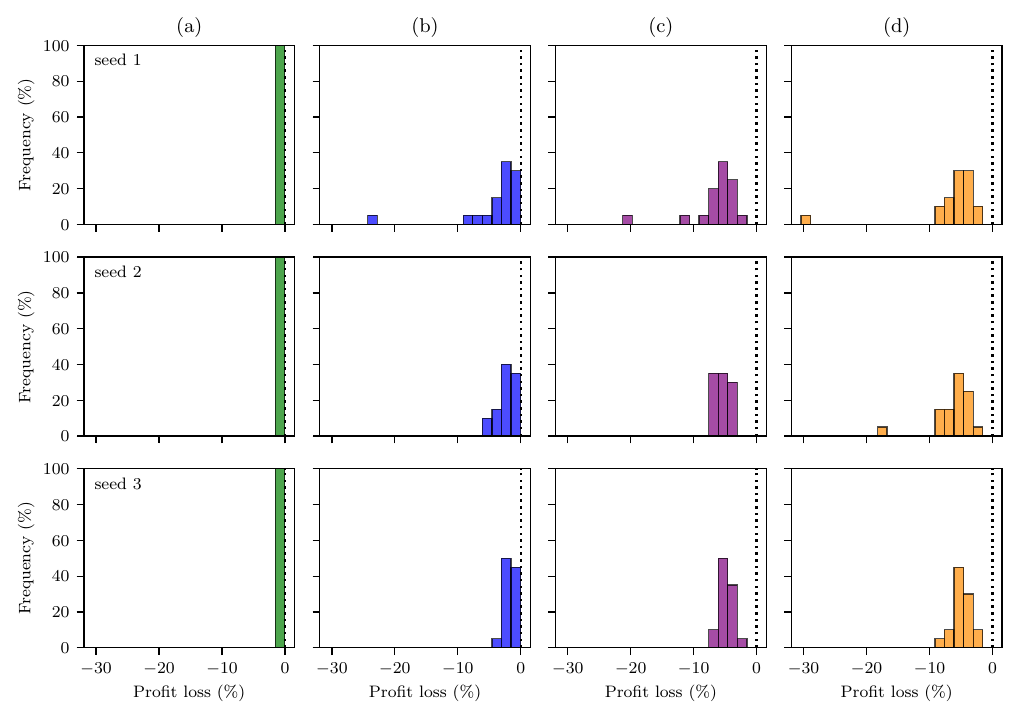}
    \caption{Profit loss $\Delta \ell$ for (a) \textit{rmeas}, (b) \textit{cmeas\_rinit},
    (c) \textit{cmeas} and (d) \textit{cmeas\_noise}. Columns are cases and rows are
    the 3 multi-start RTO seeds; each panel is the distribution over the 20 ensemble
    members, plotted as relative frequency. The dotted line marks the best achievable
    profit loss, $0\%$.}
    \label{fig:profit_loss_struc}
\end{figure*}

We next solve the actual RTO problem \cref{eq:econ_opt} and evaluate the profit loss $\Delta \ell$ 
over the 20 ensembles for each case of the structured and the black-box models 
over 3 multi-start RTO seeds. Within
each multi-start seed, the RTO problem
\cref{eq:econ_opt} for each model is initialized with the \emph{same} guess $(x_s, u_s)$, 
which is obtained by perturbing the midpoint of the training data range. This ensures 
RTO initialization is not a confounder within a seed.
As shown in \cref{fig:profit_loss_struc}, \textit{rmeas} recovers the plant 
optimum with negligible profit loss and no variability across the multi-start seeds.
This result again confirms that the 
training dataset is informative enough to identify the plant rate structure. 

Every other case, however, incurs a significant profit loss that
varies across the RTO seeds. The worst-case profit loss for structured models is just over
$30\%$ as seen from \cref{fig:profit_loss_struc} panel (d) for seed 1. Models that 
were visually indistinguishable on the validation dataset predict \emph{multiple} 
optima, even though the same multistart search returns a single optimum for the plant.
This disparity among RTO solutions is not an artifact of noise, since
apart from \textit{cmeas\_noise} all results for structured models used noise-free measurements. Nor
would training the models longer help, since the validation loss for the models has already plateaued
within the chosen epoch budget. As noted previously, a good starting point does not help either.
Even after warm-starting from the \textit{rmeas} weights and biases
that recover the plant optimum, \textit{cmeas\_rinit}
drifts and returns a near-zero loss only occasionally. 

\begin{figure}[!htbp]
    \centering
    \includegraphics[width=0.5\textwidth, page=5]{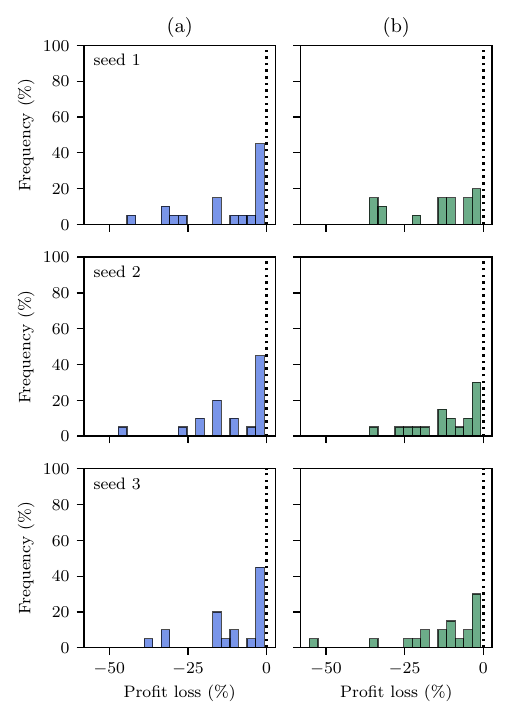}
    \caption{Profit loss $\Delta \ell$ for (a) \textit{bbox} and (b)
    \textit{bbox\_noise}. Columns are cases and rows are the 3 multi-start RTO
    seeds; each panel is the distribution over the 20 ensemble members, plotted
    as relative frequency. The dotted line marks the best achievable
    profit loss, $0\%$.}
    \label{fig:profit_loss_bbox}
\end{figure}

A natural objection, often raised in support of black-box models, is
that imposing structure itself is the culprit. 
By fixing the stoichiometry with mass balances and thermodynamics, we may have imposed
a model class too rigid for the plant, and letting the data determine the entire
dynamics might do better on RTO. We tested the claim directly with the black-box models. 
Their economic performance, however, is still worse for the actual RTO problem.
The profit loss in \cref{fig:profit_loss_bbox} exceeds that of every structured 
case, with the worst-case exceeding $50\%$ as seen from panel (b) for seed 3.

\section{Conclusions}
\label{sec:conclusions}

In this paper, we tested whether a data-driven model that predicts plant measurements accurately 
can be trusted to drive economic
decisions. Using the vinyl acetate process as a benchmark, we trained \emph{hybrid (structured)}
models, which embed a neural-network closure for reaction rates in the known mass balances
and thermodynamics, and a fully data-driven \emph{black-box} model. Within each model class, we 
considered several cases that differ in the measurements used for training and 
the noise level. We tested each case in real-time optimization.
Across the ensemble based on random initializations, 
each model and its variants cleared
the usual validation checks, with accurate fits and tight ensemble spread on the
validation dataset, yet their RTO solutions showed a wide disparity. 
In the worst case, the structured models lost $30\%$ of profit and the
black-box cases over $50\%$.
With losses this large, we may do better for economic profitability
by not using real-time optimization at all.
Additionally, under the same multistart search that returns a single optimum for the plant,
the models returned many spurious RTO solutions.

We ran these simulations under the most ideal settings we could construct. 
We took care to have the measurements at the plant optimum bracketed 
within the training dataset, ensuring that
the models are tested purely on interpolation. We trained the models on 
both noise-free and noisy measurements. 
In fact, in the 
\textit{cmeas\_rinit} case the NN weights and biases that recover the plant RTO 
landscape were handed to the training optimizer as its starting point.  
Training on concentration data drove those parameters away to a suboptimal fit 
whose RTO landscape no longer matched the one it started from. 

The observed optimizer drift is not a numerical failure. 
Stochastic optimizers such as ADAM, which are workhorses of ML/AI frameworks, advance 
without a line search, and that is precisely what lets them escape poor 
local minima. The same property, however, can let them drift away from a good 
basin. Switching to a deterministic optimizer, 
L-BFGS, for the later epochs does not repair the drift; L-BFGS performs a line search 
but by the time it takes over, the 
initial ADAM phase has already left the good basin it was initialized in, so 
L-BFGS only refines within whichever basin it inherits. A practitioner does 
not expect an optimizer to leave an optimal solution it was handed. The effect 
does not show up in the training or validation fit, but in the RTO 
decision. Nor is starting L-BFGS from the correct basin a remedy, since it 
presumes one already holds the RTO-optimal parameters.
This fragility of the standard ML training recipe is what we highlight.

This fragility 
can only worsen in the presence of process noise, unmodeled states or disturbances, or 
insufficiently exciting 
closed-loop data. We ran these investigations with the best software and 
tuned hyperparameters, and the models still failed under these \emph{ideal} conditions. 
Any industrial vendor proposing to solve a similar problem should first 
demonstrate recovery of the 
optimum on a benchmark of the type proposed here. Short of that, the attempt 
remains unreliable.

Additional regularization of the structured models could help overcome such fragility.
Future work could test whether imposing monotonicity constraints on certain inputs 
(e.g., temperature) in the NN-approximated rate laws, to match insights from 
process chemistry, improves RTO performance.
Another avenue is to impose a Lipschitz bound on the neural-network 
curvature and test whether this makes data-driven model-based RTO more robust.

\section*{Acknowledgment}
The funding for this work was provided by the Dow Chemical Company. PD acknowledges
support from the Mitsubishi Chemical Graduate Fellowship. Use was made of
computational facilities purchased with funds from the
National Science Foundation (CNS-1725797) and administered by the Center for
Scientific Computing (CSC). The CSC is supported by the California NanoSystems
Institute and the Materials Research Science and Engineering Center (MRSEC;
NSF DMR 2308708) at UC Santa Barbara.

\section*{Data availability}
The code and simulated data used to reproduce all results are available at
\url{https://dakeprithvi.github.io/2026c_structure_id_plant} and 
\cite{dake:bindlish:rawlings:2026a}.

\appendix
\crefalias{section}{appendix}
\makeatletter
\gdef\thesection{\@Alph\c@section}%
\def\@seccntformat#1{\appendixname~\csname the#1\endcsname\@seccntDot\hskip 0.5em}%
\makeatother
\section{Process units}
\label{app:process_units}
Every separator and splitter acts as a linear split: an inlet molar flow
$N_j^{\rm in}$ of species $j$ divides into a split-off stream and a remainder,
\begin{equation*}
    N_j^{\rm split} = \delta_j\, N_j^{\rm in}, \qquad
    N_j^{\rm rem}   = (1-\delta_j)\, N_j^{\rm in}, \qquad j \in J,
\end{equation*}
where $J = \{\Rm{Ee, Ea, A, V, W, O, C}\}$ denotes ethylene, ethane, acetic acid, 
vinyl acetate, water, oxygen, and carbon dioxide. \Cref{tab:units} lists
each unit, its inlet/outlet streams, and its split fraction. Each unit carries
its own symbol for $\delta_j$ so the flowsheet equations stay unambiguous,
$\phi_j$ for the flash drum, $\gamma_j$ for the distillation column, $\mu_j$ for
the absorber, and the recycle fractions $\alpha$ and $\beta$ for splitters I and
II.
\begin{table}[h]
    \caption{Linear split for each separation unit and splitter. Stream
    superscripts follow the flowsheet; the last column gives each unit's split
    fraction, the fraction of $N_j^{\rm in}$ routed to the split-off stream.}
    \centering
    \begin{tabularx}{\linewidth}{@{}l c c >{\raggedright\arraybackslash}X@{}}
    \toprule
    \textbf{Unit} & \textbf{In} & \textbf{Split/Rem} & \textbf{Split fraction} \\
    \midrule
    Flash drum & $N^3$ & $N^5/N^4$    & $\phi_j=1$: Ee,Ea,O,C; $0$: A,V,W \\
    Distillation column   & $N^4$ & $N^8/N^{10}$ & $\gamma_j=1$: A; $0$: otherwise \\
    Absorber              & $N^5$ & $N^{11}/N^6$ & $\mu_j=0.03$: C; $0$: otherwise \\
    Splitter I            & $N^6$ & $N^7/N^9$    & $\alpha$ (all $j$) \\
    Splitter II           & $N^8$ & $N^{13}/N^{12}$ & $\beta$ (all $j$) \\
    \bottomrule
    \end{tabularx}
    \label{tab:units}
\end{table}
\section{Parameter values}
\label{app:parameter_values}
The price-vector is given as,
$(p_1, p_2, p_3, p_{4,\Rm{Ee}}, p_{4,\Rm{O}}, p_{4,\Rm{A}}, p_5, p_6)=$ 
(300, 33, 50, 100, 50, 50, 30, 0) \unitfrac{\$}{mol} 
\begin{table}[h]
    \caption{Parameter values used for the VAc model.}
    \centering
    \begin{tabularx}{\linewidth}{*{3}{>{\centering\arraybackslash}X}}
    \toprule
    \textbf{Parameter} & \textbf{Value} & \textbf{Units}  \\
    \midrule
    $k_1$ & $1.312\times10^{-2}$ & $\unitfrac{mol}{kg\,s}\left(\unitfrac{m^3}{mol}\right)^{3.2}$ \\
    $E_1$ & 40500 & J/mol \\
    $k_2$ & $2.363$ & $\unitfrac{mol}{kg\,s}\left(\unitfrac{m^3}{mol}\right)^{2}$ \\
    $E_2$ & 52500 & J/mol \\
    R & 8.314 & \unitfrac{J}{mol K} \\
    $P$ & 882.529 & kPa \\
    $V_R$ & 6.68 & $\Rm{m}^3$ \\
    $\epsilon$ & 0.8 & - \\
    $\rho_c$ & 385 & $\unitfrac{kg}{m^3}$ \\
    $\lambda$ & 0.001 & - \\
    $\underline{u}$ &
    (10.88, 7.8, 12.01, 0.4, 0.6, 384.39) & (\unitfrac{mol}{s}, \unitfrac{mol}{s}, \unitfrac{mol}{s}, -, -, $\Rm{K}$)\\
    $\bar{u}$ &
    (20.20, 14.49, 22.30, 0.74, 0.6, 469.80) & (\unitfrac{mol}{s}, \unitfrac{mol}{s}, \unitfrac{mol}{s}, -, -, $\Rm{K}$)\\
    $s_u$ &
    (13.85, 13.85, 13.08, 1, 1, 423.15) & (\unitfrac{mol}{s}, \unitfrac{mol}{s}, \unitfrac{mol}{s}, -, -, $\Rm{K}$)\\
    \bottomrule
    \end{tabularx}
    \label{tab:training_summary}
\end{table}

\bibliographystyle{abbrvnat}
\bibliography{paper}

\end{document}


\maketitle

\section{Learning-rate sweep and optimizer comparison}
\label{sec:lr_sweep}

For reference, we restate the training problems whose learning rate is swept
here. The concentration-measurement cases \textit{cmeas\_rinit}, \textit{cmeas},
and \textit{cmeas\_noise} solve the \emph{structured} training problem, Eq.~(9)
of the main paper,
\begin{align}
	\Theta^\star_\Rm{NN} &= \argmin_{\Theta_\Rm{NN}}
		 \left( \frac{1}{N_{tr}N_t} \sum_{i=1}^{N_{tr}} \sum_{k=0}^{N_t}
          \norm{ \frac{y_i(k) - \yhat_i(k)}{\sigma_y} }^2 \right)^{1/2} \notag \\
		\textnormal{s.t.} \quad \dot{\hat{x}}_{i} &= f(\hat{x}_{i}, u_i; \Theta_\Rm{NN}) \notag \\
		\yhat_i &= h(\hat{x}_{i}, u_i) \notag \\
        \hat{x}_i(0) &= f_0(y_i(0))
        \label{eq:supp_struc_train}
\end{align}
while the two black-box cases \textit{bbox} and \textit{bbox\_noise} solve the
\emph{black-box} training problem, Eq.~(11) of the main paper,
\begin{align}
	\theta^\star_{f_{\Rm{NN}}} &= \argmin_{\theta_{f_{\Rm{NN}}}}
		 \left( \frac{1}{N_{tr} N_t} \sum_{i=1}^{N_{tr}} \sum_{k=0}^{N_t}
          \norm{ \frac{y_i(k) - \yhat_i(k)}{\sigma_y} }^2 \right)^{1/2} \notag \\
		\textnormal{s.t.} \quad \dot{\hat{x}}_{i} &= f_{\Rm{NN}}(\hat{x}_{i}, u_i; \theta_{f_{\Rm{NN}}}) \notag \\
		\yhat_i &= h(\hat{x}_{i}, u_i) \notag \\
        \hat{x}_i(0) &= f_0(y_i(0))
        \label{eq:supp_blackbox_nn}
\end{align}
The symbols are defined in the main paper. Every sweep in this section is
performed at the fixed batch count $N_B = 5$ used throughout the paper (the
$N_{tr}$ training trajectories are partitioned into $N_B$ batches of size
$N_{tr}/N_B$ during the stochastic ADAM phase); only the learning rate is
varied.

\begin{figure}[htbp]
    \centering
    \includegraphics[width=\textwidth, page=1]{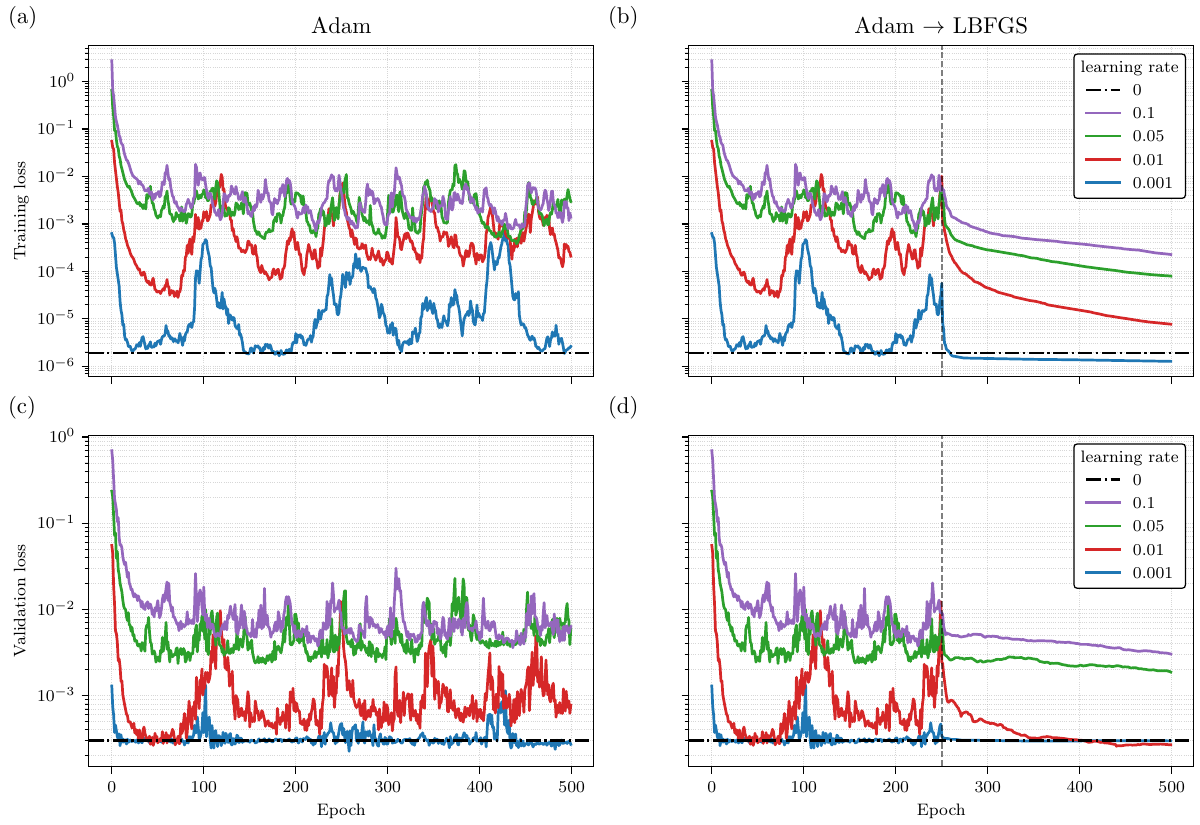}
    \caption{\textit{cmeas\_rinit}: drift away from the initialization. The
    model is initialized at the weights at which it recovers the plant optimum,
    marked by the dash-dotted line (learning rate $0$, no training step taken).
    Under ADAM (left) the loss moves away from that line as training proceeds,
    so the optimizer leaves the basin it was handed. After the switch to L-BFGS
    at epoch 250 (right) the loss returns toward the initialization, so L-BFGS
    recovers part of what ADAM gave up. As the learning rate decreases, ADAM
    takes smaller steps and the curves stay at the initialization; this
    preserves the optimum only because it was supplied, not because training
    located it.}
    \label{fig:lr_cmeas_rinit}
\end{figure}

\begin{figure}[htbp]
    \centering
    \includegraphics[width=\textwidth, page=2]{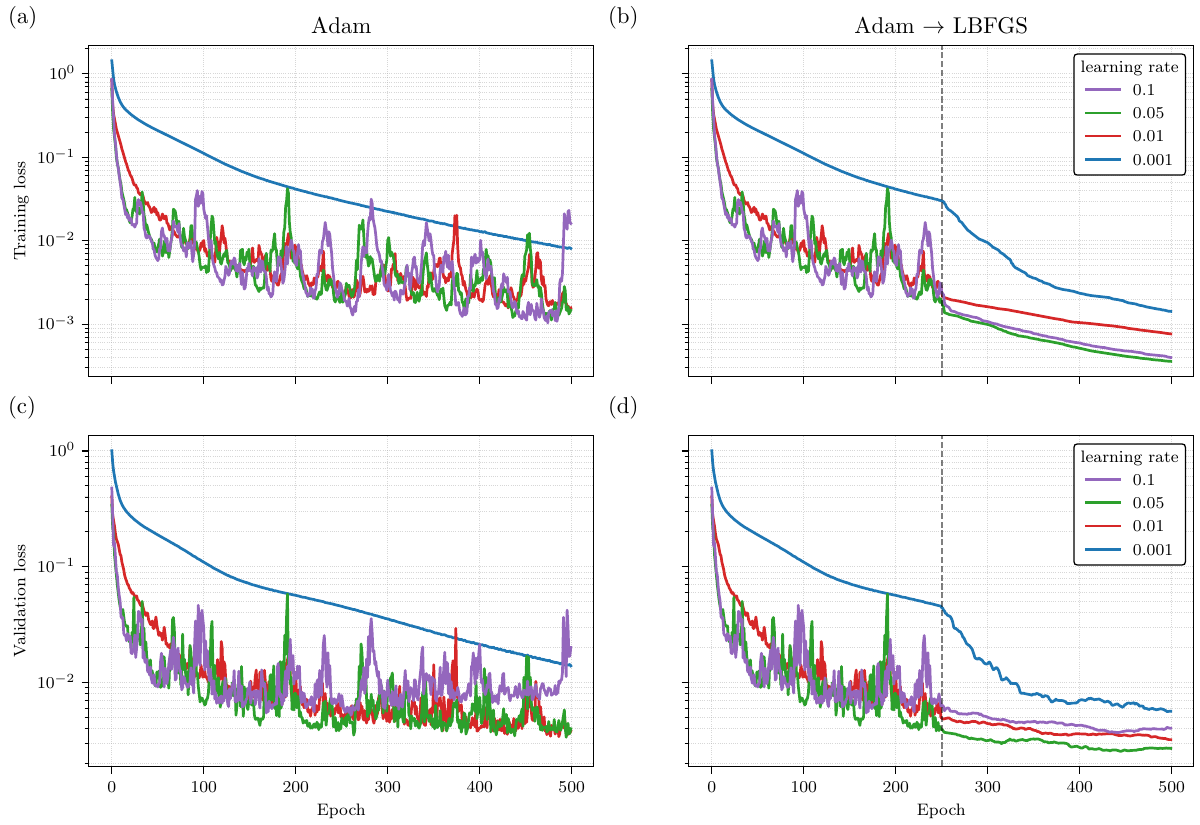}
    \caption{Learning-rate sweep for \textit{cmeas}. The chosen
    learning rate is $0.05$.}
    \label{fig:lr_cmeas}
\end{figure}

\begin{figure}[htbp]
    \centering
    \includegraphics[width=\textwidth, page=3]{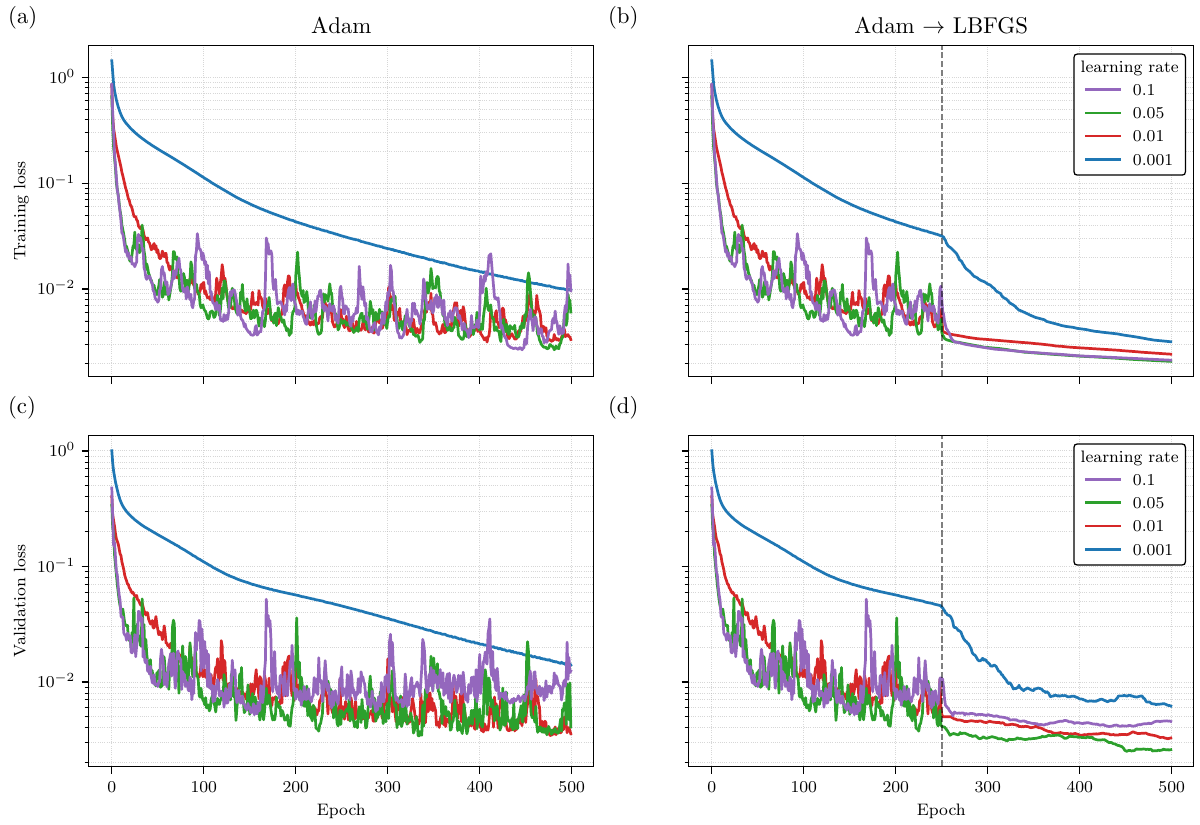}
    \caption{Learning-rate sweep for \textit{cmeas\_noise}. The chosen
    learning rate is $0.05$.}
    \label{fig:lr_cmeas_noise}
\end{figure}

\begin{figure}[htbp]
    \centering
    \includegraphics[width=\textwidth, page=4]{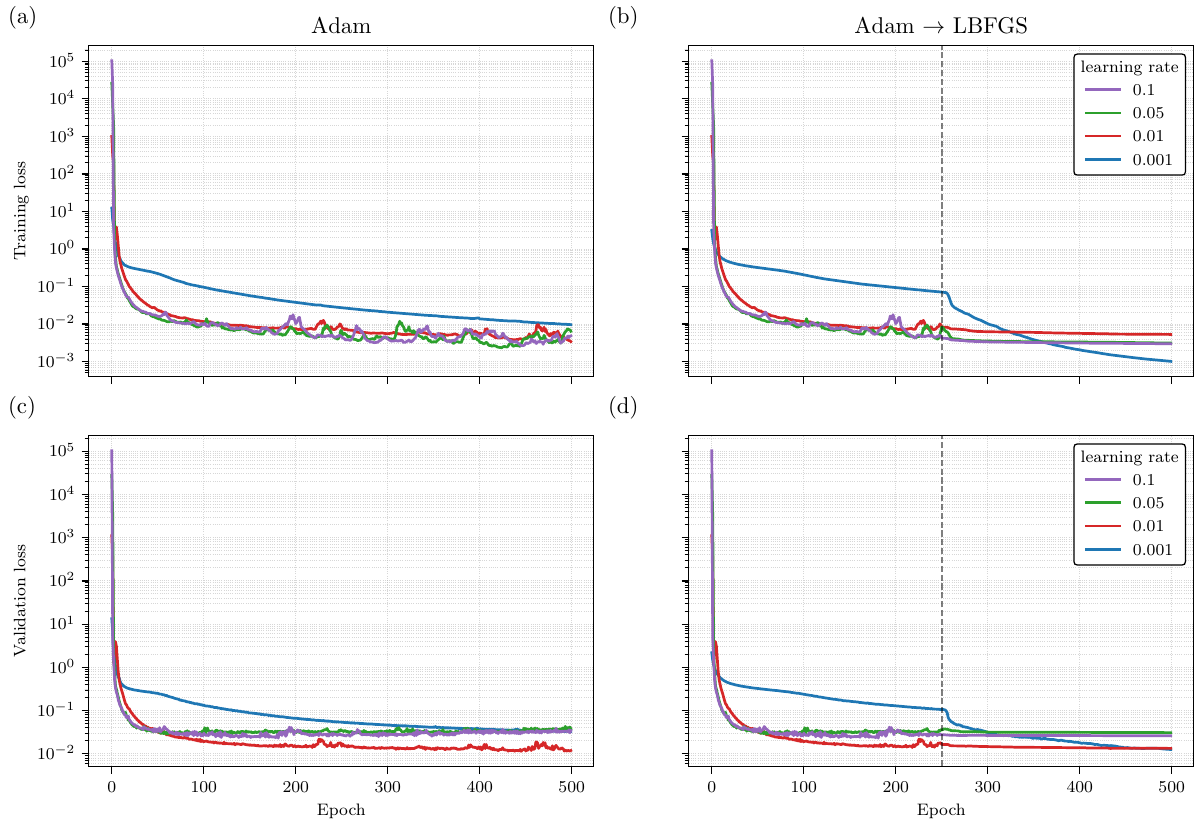}
    \caption{Learning-rate sweep for \textit{bbox}. The chosen
    learning rate is $0.01$.}
    \label{fig:lr_bbox}
\end{figure}

\begin{figure}[htbp]
    \centering
    \includegraphics[width=\textwidth, page=5]{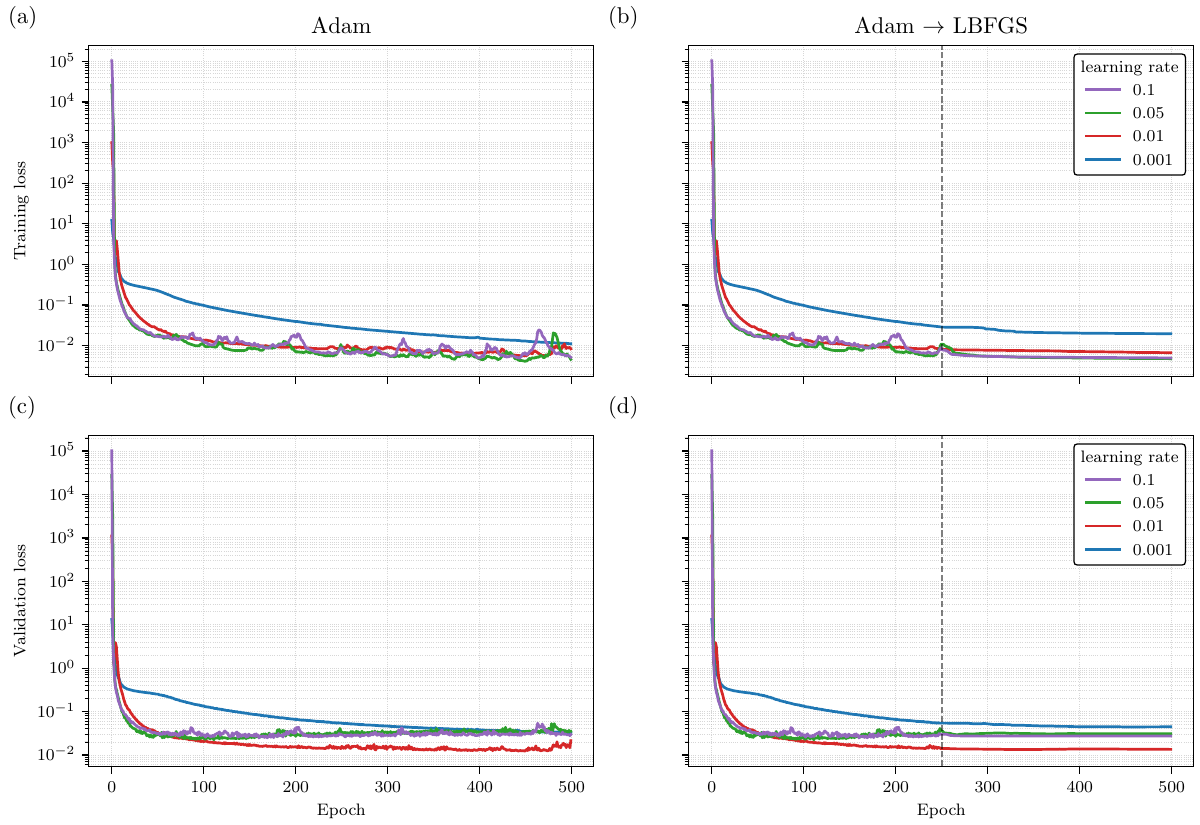}
    \caption{Learning-rate sweep for \textit{bbox\_noise}. The chosen
    learning rate is $0.01$.}
    \label{fig:lr_bbox_noise}
\end{figure}

Figures~\ref{fig:lr_cmeas_rinit}--\ref{fig:lr_bbox_noise} show the
learning-rate sweep for each case, and every figure has the same four-panel
layout. The top row is the training loss and the bottom row the validation loss,
both on a logarithmic scale; the left column is ADAM alone and the right column
is ADAM followed by L-BFGS. Panel (a) is therefore the training loss under ADAM,
(b) the training loss under ADAM and then L-BFGS, (c) the validation loss under
ADAM, and (d) the validation loss under ADAM and then L-BFGS. Within each panel we
overlay one curve per learning rate across the full epoch budget, so a single
panel shows both the learning-rate sweep and whether the loss has plateaued
within that budget. In the right column, panels (b) and (d), the dashed vertical
line at epoch 250 marks the switch from Adam to L-BFGS.

For \textit{cmeas} and \textit{cmeas\_noise}, and for the two black-box cases, we
select the learning rate that gives the lowest validation loss in panel (d) and
confirm from the same panel that the validation loss has plateaued within the
epoch budget.

\textit{cmeas\_rinit} is read differently. It is initialized at the weights at
which the model recovers the plant optimum, so no learning rate is selected from
it. In the paper, however we show results for the same learning rate as chosen for 
\textit{cmeas} and \textit{cmeas\_noise}, i.e., 0.05. 
The dash-dotted horizontal line, labelled learning rate $0$, is the loss at
that initialization before any training step is taken.

%% file: abstract.tex
\begingroup%
  \makeatletter%
  \providecommand\color[2][]{%
    \errmessage{(Inkscape) Color is used for the text in Inkscape, but the package 'color.sty' is not loaded}%
    \renewcommand\color[2][]{}%
  }%
  \providecommand\transparent[1]{%
    \errmessage{(Inkscape) Transparency is used (non-zero) for the text in Inkscape, but the package 'transparent.sty' is not loaded}%
    \renewcommand\transparent[1]{}%
  }%
  \providecommand\rotatebox[2]{#2}%
  \newcommand*\fsize{\dimexpr\f@size pt\relax}%
  \newcommand*\lineheight[1]{\fontsize{\fsize}{#1\fsize}\selectfont}%
  \ifx\svgwidth\undefined%
    \setlength{\unitlength}{529.93094377bp}%
    \ifx\svgscale\undefined%
      \relax%
    \else%
      \setlength{\unitlength}{\unitlength * \real{\svgscale}}%
    \fi%
  \else%
    \setlength{\unitlength}{\svgwidth}%
  \fi%
  \global\let\svgwidth\undefined%
  \global\let\svgscale\undefined%
  \makeatother%
  \begin{picture}(1,0.84631526)%
    \lineheight{1}%
    \setlength\tabcolsep{0pt}%
    \put(0,0){\includegraphics[width=\unitlength,page=1]{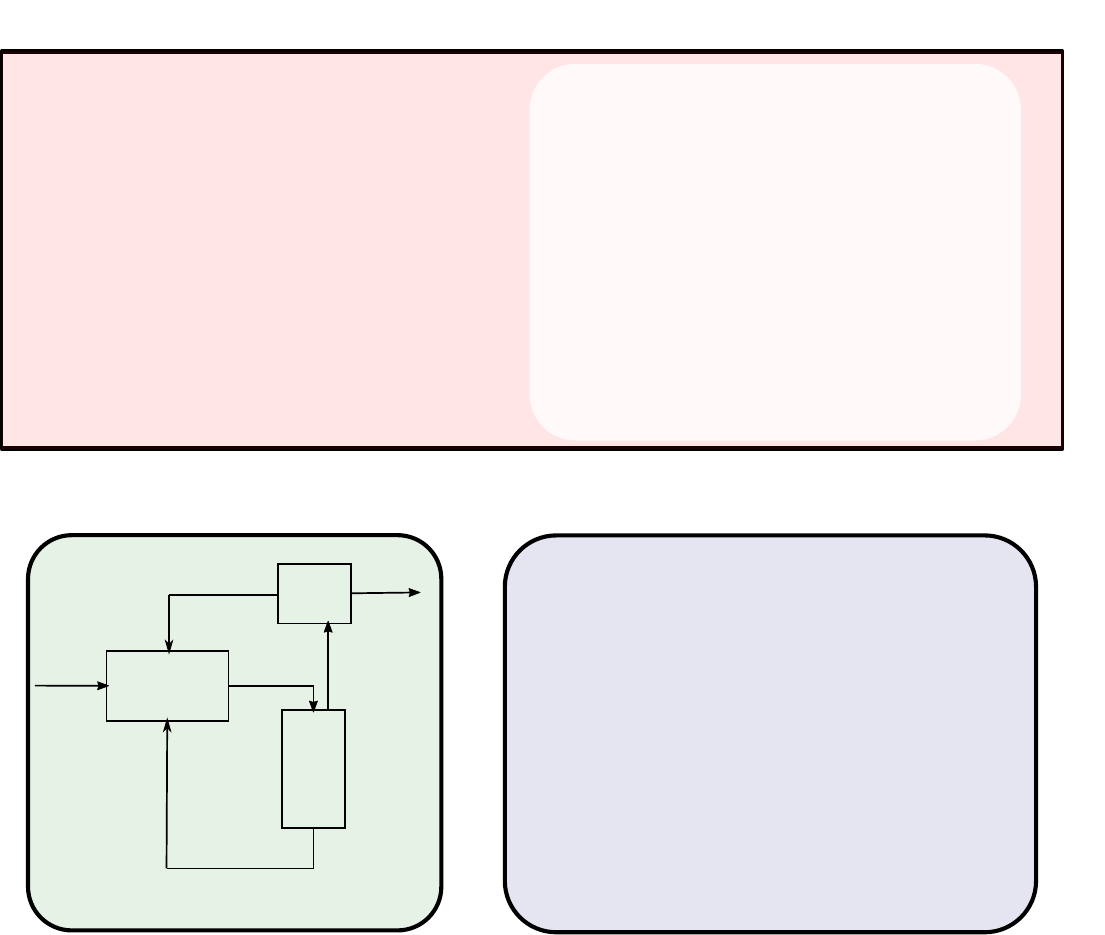}}%
    \put(0.13750739,0.21276031){\makebox(0,0)[lt]{\lineheight{1.25}\smash{\begin{tabular}[t]{l}R\end{tabular}}}}%
    \put(0.27467156,0.29589763){\makebox(0,0)[lt]{\lineheight{1.25}\smash{\begin{tabular}[t]{l}S\end{tabular}}}}%
    \put(0.27020618,0.13704247){\makebox(0,0)[lt]{\lineheight{1.25}\smash{\begin{tabular}[t]{l}D\end{tabular}}}}%
    \put(0,0){\includegraphics[width=\unitlength,page=2]{abstract-fig.pdf}}%
    \put(0.07982908,0.37574135){\makebox(0,0)[lt]{\lineheight{1.25}\smash{\begin{tabular}[t]{l}Chemical Plant\end{tabular}}}}%
    \put(0.03606344,0.3078006){\makebox(0,0)[lt]{\lineheight{1.25}\smash{\begin{tabular}[t]{l}(A)\end{tabular}}}}%
    \put(0.03546671,0.74024997){\makebox(0,0)[lt]{\lineheight{1.25}\smash{\begin{tabular}[t]{l}(B)\end{tabular}}}}%
    \put(0,0){\includegraphics[width=\unitlength,page=3]{abstract-fig.pdf}}%
    \put(0.11744783,0.53005579){\rotatebox{89.40488}{\makebox(0,0)[lt]{\lineheight{1.25}\smash{\begin{tabular}[t]{l}Plant Data\end{tabular}}}}}%
    \put(0,0){\includegraphics[width=\unitlength,page=4]{abstract-fig.pdf}}%
    \put(0.24156364,0.45124881){\makebox(0,0)[lt]{\lineheight{1.25}\smash{\begin{tabular}[t]{l}t\end{tabular}}}}%
    \put(0,0){\includegraphics[width=\unitlength,page=5]{abstract-fig.pdf}}%
    \put(0.4046343,0.81862334){\makebox(0,0)[lt]{\lineheight{1.25}\smash{\begin{tabular}[t]{l}Training\end{tabular}}}}%
    \put(0.58301709,0.75242741){\makebox(0,0)[lt]{\lineheight{1.25}\smash{\begin{tabular}[t]{l}Structured model\end{tabular}}}}%
    \put(0.51836039,0.67031117){\makebox(0,0)[lt]{\lineheight{1.25}\smash{\begin{tabular}[t]{l}$\dfrac{dc_j}{dt} =\dfrac{c_{jf}-c_j}{\tau}+\sum_i\nu_{ij} $\end{tabular}}}}%
    \put(0.60965886,0.51070735){\makebox(0,0)[lt]{\lineheight{1.25}\smash{\begin{tabular}[t]{l}$\dfrac{dc_j}{dt} = $\end{tabular}}}}%
    \put(0.60559759,0.59945963){\makebox(0,0)[lt]{\lineheight{1.25}\smash{\begin{tabular}[t]{l}Black-box model\end{tabular}}}}%
    \put(0.10909796,0.02168609){\makebox(0,0)[lt]{\lineheight{1.25}\smash{\begin{tabular}[t]{l}ML/AI closure\end{tabular}}}}%
    \put(0.37004351,0.59389858){\makebox(0,0)[lt]{\lineheight{1.25}\smash{\begin{tabular}[t]{l}u\end{tabular}}}}%
    \put(0.37019638,0.74352892){\makebox(0,0)[lt]{\lineheight{1.25}\smash{\begin{tabular}[t]{l}y\end{tabular}}}}%
    \put(0,0){\includegraphics[width=\unitlength,page=6]{abstract-fig.pdf}}%
    \put(0.59618576,0.30003057){\color[rgb]{0.10196078,0.10196078,0.10196078}\makebox(0,0)[lt]{\lineheight{1.25}\smash{\begin{tabular}[t]{l}$u_s^* = \text{argmax} \ \ell(x_s, u_s)$\end{tabular}}}}%
    \put(0.65288756,0.2712483){\color[rgb]{0.10196078,0.10196078,0.10196078}\makebox(0,0)[lt]{\lineheight{1.25}\smash{\begin{tabular}[t]{l}$0=f(x_s, u_s)$\end{tabular}}}}%
    \put(0.6266139,0.02487639){\makebox(0,0)[lt]{\lineheight{1.25}\smash{\begin{tabular}[t]{l}Profit loss\end{tabular}}}}%
    \put(0,0){\includegraphics[width=\unitlength,page=7]{abstract-fig.pdf}}%
    \put(0.50286241,0.09759466){\rotatebox{89.40488}{\makebox(0,0)[lt]{\lineheight{1.25}\smash{\begin{tabular}[t]{l}Counts\end{tabular}}}}}%
    \put(0.52133587,0.22781596){\color[rgb]{0.10196078,0.10196078,0.10196078}\makebox(0,0)[lt]{\lineheight{1.25}\smash{\begin{tabular}[t]{l}$<30\%$\end{tabular}}}}%
    \put(0.73691348,0.22782363){\color[rgb]{0.10196078,0.10196078,0.10196078}\makebox(0,0)[lt]{\lineheight{1.25}\smash{\begin{tabular}[t]{l}$<100\%$\end{tabular}}}}%
    \put(0,0){\includegraphics[width=\unitlength,page=8]{abstract-fig.pdf}}%
    \put(0.50254926,0.37550575){\makebox(0,0)[lt]{\lineheight{1.25}\smash{\begin{tabular}[t]{l}Economic Optimization\end{tabular}}}}%
    \put(0.5840962,0.17946116){\makebox(0,0)[lt]{\lineheight{1.25}\smash{\begin{tabular}[t]{l}\Huge{\textcolor{black}{RELIABLE?}}\end{tabular}}}}%
    \put(0.4625298,0.3095978){\makebox(0,0)[lt]{\lineheight{1.25}\smash{\begin{tabular}[t]{l}(C)\end{tabular}}}}%
    \put(0.39498348,0.36418351){\color[rgb]{0.4,0.4,0.4}\makebox(0,0)[lt]{\lineheight{1.25}\smash{\begin{tabular}[t]{l}RTO\end{tabular}}}}%
  \end{picture}%
\endgroup%

%% file: Economics.tex
\begingroup%
  \makeatletter%
  \providecommand\color[2][]{%
    \errmessage{(Inkscape) Color is used for the text in Inkscape, but the package 'color.sty' is not loaded}%
    \renewcommand\color[2][]{}%
  }%
  \providecommand\transparent[1]{%
    \errmessage{(Inkscape) Transparency is used (non-zero) for the text in Inkscape, but the package 'transparent.sty' is not loaded}%
    \renewcommand\transparent[1]{}%
  }%
  \providecommand\rotatebox[2]{#2}%
  \newcommand*\fsize{\dimexpr\f@size pt\relax}%
  \newcommand*\lineheight[1]{\fontsize{\fsize}{#1\fsize}\selectfont}%
  \ifx\svgwidth\undefined%
    \setlength{\unitlength}{625.03318486bp}%
    \ifx\svgscale\undefined%
      \relax%
    \else%
      \setlength{\unitlength}{\unitlength * \real{\svgscale}}%
    \fi%
  \else%
    \setlength{\unitlength}{\svgwidth}%
  \fi%
  \global\let\svgwidth\undefined%
  \global\let\svgscale\undefined%
  \makeatother%
  \begin{picture}(1,0.71659533)%
    \lineheight{1}%
    \setlength\tabcolsep{0pt}%
    \put(0.24574502,0.61422266){\color[rgb]{0,0,1}\makebox(0,0)[lt]{\lineheight{1.25}\smash{\begin{tabular}[t]{l}7\end{tabular}}}}%
    \put(0.04249832,0.45273059){\color[rgb]{1,0,0}\makebox(0,0)[lt]{\lineheight{1.25}\smash{\begin{tabular}[t]{l}\$\end{tabular}}}}%
    \put(0.23141832,0.65259884){\color[rgb]{1,0,0}\makebox(0,0)[lt]{\lineheight{1.25}\smash{\begin{tabular}[t]{l}\$\end{tabular}}}}%
    \put(0.25539334,0.06828554){\color[rgb]{1,0,0}\makebox(0,0)[lt]{\lineheight{1.25}\smash{\begin{tabular}[t]{l}\$\end{tabular}}}}%
    \put(0.90215071,0.21602564){\color[rgb]{0,1,0}\makebox(0,0)[lt]{\lineheight{1.25}\smash{\begin{tabular}[t]{l}\$\end{tabular}}}}%
    \put(0,0){\includegraphics[width=\unitlength,page=1]{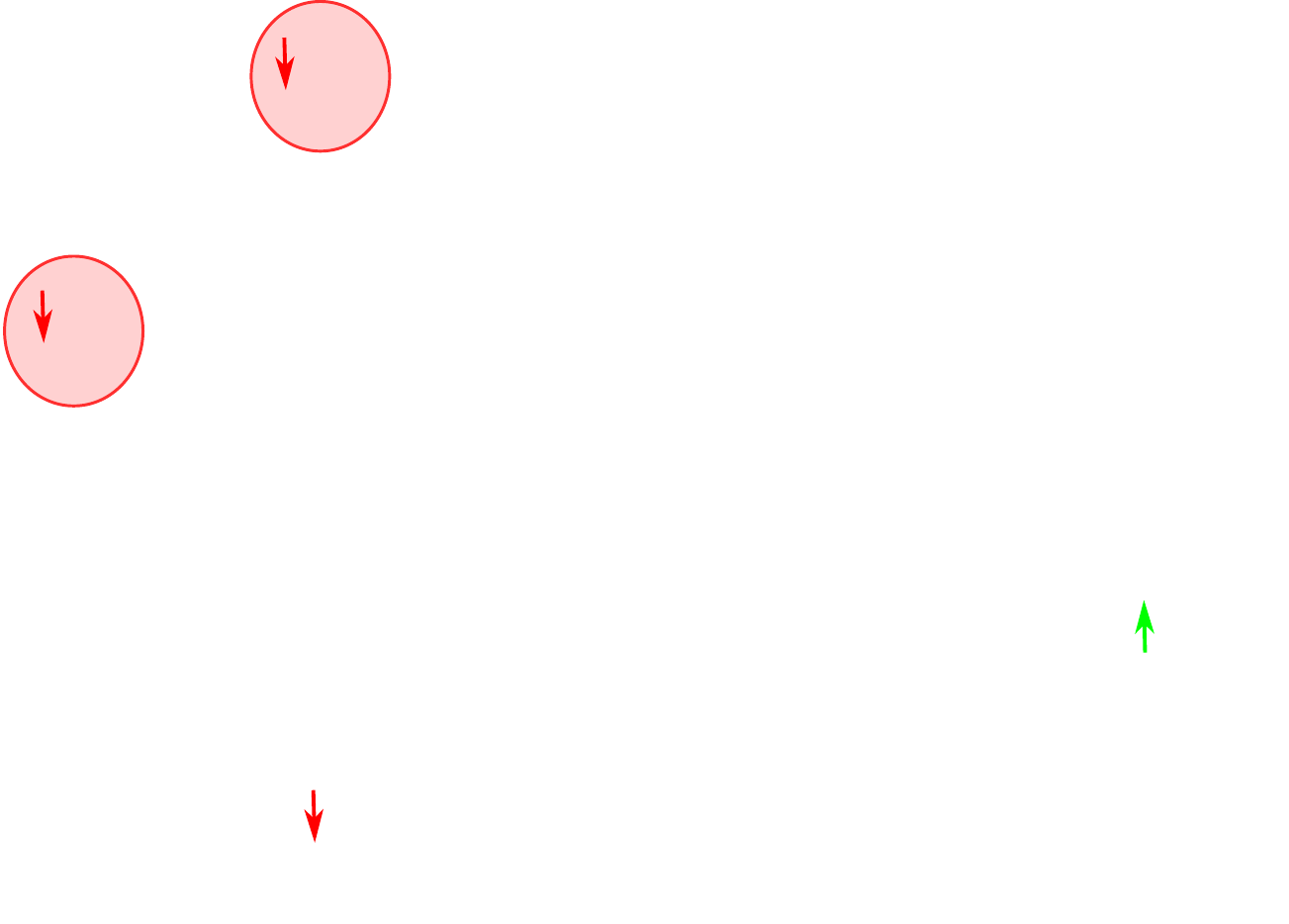}}%
    \put(0.90125289,0.64844586){\color[rgb]{1,0,0}\makebox(0,0)[lt]{\lineheight{1.25}\smash{\begin{tabular}[t]{l}\$\end{tabular}}}}%
    \put(0,0){\includegraphics[width=\unitlength,page=2]{Economics-fig.pdf}}%
    \put(0.89770181,0.07610635){\color[rgb]{1,0,0}\makebox(0,0)[lt]{\lineheight{1.25}\smash{\begin{tabular}[t]{l}\$\end{tabular}}}}%
    \put(0,0){\includegraphics[width=\unitlength,page=3]{Economics-fig.pdf}}%
    \put(0.49014237,0.63702391){\color[rgb]{1,0,0}\makebox(0,0)[lt]{\lineheight{1.25}\smash{\begin{tabular}[t]{l}Gas recycle $\alpha$\end{tabular}}}}%
    \put(0,0){\includegraphics[width=\unitlength,page=4]{Economics-fig.pdf}}%
    \put(0.43993125,0.32388811){\makebox(0,0)[t]{\lineheight{1.25}\smash{\begin{tabular}[t]{c}Reactor\end{tabular}}}}%
    \put(0.01150823,0.36244863){\makebox(0,0)[t]{\lineheight{1.25}\smash{\begin{tabular}[t]{c}$N^1_{\mathrm{Ee}}$\end{tabular}}}}%
    \put(0.00410889,0.32824134){\makebox(0,0)[t]{\lineheight{1.25}\smash{\begin{tabular}[t]{c}$N^1_{\mathrm{A}}$\end{tabular}}}}%
    \put(0.00574874,0.29334165){\makebox(0,0)[t]{\lineheight{1.25}\smash{\begin{tabular}[t]{c}$N^1_{\mathrm{O}}$\end{tabular}}}}%
    \put(0,0){\includegraphics[width=\unitlength,page=5]{Economics-fig.pdf}}%
    \put(0.75208786,0.32839275){\makebox(0,0)[t]{\lineheight{1.25}\smash{\begin{tabular}[t]{c}Flash\end{tabular}}}}%
    \put(0,0){\includegraphics[width=\unitlength,page=6]{Economics-fig.pdf}}%
    \put(0.75573699,0.17296803){\makebox(0,0)[t]{\lineheight{1.25}\smash{\begin{tabular}[t]{c}Distillation\\ column\end{tabular}}}}%
    \put(0,0){\includegraphics[width=\unitlength,page=7]{Economics-fig.pdf}}%
    \put(0.7531137,0.47618773){\makebox(0,0)[t]{\lineheight{1.25}\smash{\begin{tabular}[t]{c}Absorber\end{tabular}}}}%
    \put(0,0){\includegraphics[width=\unitlength,page=8]{Economics-fig.pdf}}%
    \put(0.1859827,0.32473915){\makebox(0,0)[t]{\lineheight{1.25}\smash{\begin{tabular}[t]{c}Mixer\end{tabular}}}}%
    \put(0,0){\includegraphics[width=\unitlength,page=9]{Economics-fig.pdf}}%
    \put(0.75217519,0.59680204){\makebox(0,0)[t]{\lineheight{1.25}\smash{\begin{tabular}[t]{c}Splitter I\end{tabular}}}}%
    \put(0,0){\includegraphics[width=\unitlength,page=10]{Economics-fig.pdf}}%
    \put(0.04737457,0.40788342){\color[rgb]{0,0,1}\makebox(0,0)[lt]{\lineheight{1.25}\smash{\begin{tabular}[t]{l}1\end{tabular}}}}%
    \put(0.27594682,0.3411403){\color[rgb]{0,0,1}\makebox(0,0)[lt]{\lineheight{1.25}\smash{\begin{tabular}[t]{l}2\end{tabular}}}}%
    \put(0.75875065,0.0944368){\color[rgb]{0,0,1}\makebox(0,0)[lt]{\lineheight{1.25}\smash{\begin{tabular}[t]{l}8\end{tabular}}}}%
    \put(0.61692126,0.34150042){\color[rgb]{0,0,1}\makebox(0,0)[lt]{\lineheight{1.25}\smash{\begin{tabular}[t]{l}3\end{tabular}}}}%
    \put(0.7577491,0.24366197){\color[rgb]{0,0,1}\makebox(0,0)[lt]{\lineheight{1.25}\smash{\begin{tabular}[t]{l}4\end{tabular}}}}%
    \put(0.89996878,0.17516639){\color[rgb]{0,0,1}\makebox(0,0)[lt]{\lineheight{1.25}\smash{\begin{tabular}[t]{l}10\end{tabular}}}}%
    \put(0.90770939,0.495131){\color[rgb]{0,0,1}\makebox(0,0)[lt]{\lineheight{1.25}\smash{\begin{tabular}[t]{l}11\end{tabular}}}}%
    \put(0.91295501,0.61514903){\color[rgb]{0,0,1}\makebox(0,0)[lt]{\lineheight{1.25}\smash{\begin{tabular}[t]{l}9\end{tabular}}}}%
    \put(0.75595394,0.53041626){\color[rgb]{0,0,1}\makebox(0,0)[lt]{\lineheight{1.25}\smash{\begin{tabular}[t]{l}6\end{tabular}}}}%
    \put(0.75562952,0.40727441){\color[rgb]{0,0,1}\makebox(0,0)[lt]{\lineheight{1.25}\smash{\begin{tabular}[t]{l}5\end{tabular}}}}%
    \put(0,0){\includegraphics[width=\unitlength,page=11]{Economics-fig.pdf}}%
    \put(0.89884827,0.04317126){\color[rgb]{0,0,1}\makebox(0,0)[lt]{\lineheight{1.25}\smash{\begin{tabular}[t]{l}12\end{tabular}}}}%
    \put(0.75573699,0.02417793){\makebox(0,0)[t]{\lineheight{1.25}\smash{\begin{tabular}[t]{c}Splitter II\end{tabular}}}}%
    \put(0,0){\includegraphics[width=\unitlength,page=12]{Economics-fig.pdf}}%
    \put(0.23648741,0.03837158){\color[rgb]{0,0,1}\makebox(0,0)[lt]{\lineheight{1.25}\smash{\begin{tabular}[t]{l}13\end{tabular}}}}%
    \put(0.47587599,0.05625596){\color[rgb]{1,0,0}\makebox(0,0)[lt]{\lineheight{1.25}\smash{\begin{tabular}[t]{l}Liquid recycle $\beta$\end{tabular}}}}%
    \put(0,0){\includegraphics[width=\unitlength,page=13]{Economics-fig.pdf}}%
    \put(0.62421448,0.31137412){\makebox(0,0)[t]{\lineheight{1.25}\smash{\begin{tabular}[t]{c}$c_{\mathrm{Ee}}$\end{tabular}}}}%
    \put(0.62710828,0.28415214){\makebox(0,0)[t]{\lineheight{1.25}\smash{\begin{tabular}[t]{c}$c_{\mathrm{Ea}}$\end{tabular}}}}%
    \put(0.62039628,0.25157724){\makebox(0,0)[t]{\lineheight{1.25}\smash{\begin{tabular}[t]{c}$c_{\mathrm{A}}$\end{tabular}}}}%
    \put(0.62371367,0.21937728){\makebox(0,0)[t]{\lineheight{1.25}\smash{\begin{tabular}[t]{c}$c_{\mathrm{W}}$\end{tabular}}}}%
    \put(0.61941438,0.1892652){\makebox(0,0)[t]{\lineheight{1.25}\smash{\begin{tabular}[t]{c}$c_{\mathrm{V}}$\end{tabular}}}}%
    \put(0.62083426,0.15583099){\makebox(0,0)[t]{\lineheight{1.25}\smash{\begin{tabular}[t]{c}$c_{\mathrm{O}}$\end{tabular}}}}%
    \put(0.61756358,0.12353335){\makebox(0,0)[t]{\lineheight{1.25}\smash{\begin{tabular}[t]{c}$c_{\mathrm{C}}$\end{tabular}}}}%
    \put(0,0){\includegraphics[width=\unitlength,page=14]{Economics-fig.pdf}}%
    \put(0.26347431,0.20448282){\color[rgb]{0.10196078,0.10196078,0.10196078}\makebox(0,0)[lt]{\lineheight{1.25}\smash{\begin{tabular}[t]{l}TC\end{tabular}}}}%
  \end{picture}%
\endgroup%

%% file: paper.bib
@article{naysmith:douglas:1995,
  author =        {Naysmith, Matthew R and Douglas, Peter L},
  journal =       {},
  number =        {2},
  pages =         {67--87},
  publisher =     {Wiley Online Library},
  title =         {Review of real time optimization in the chemical
                   process industries},
  volume =        {3},
  year =          {1995},
}

@article{darby:nikolaou:jones:2011,
  author =        {Darby, Mark L. and Nikolaou, Michael and Jones, James and
                   Nicholson, Doug},
  journal =       {J. Proc. Cont.},
  number =        {6},
  pages =         {874--884},
  title =         {{RTO}: {An} overview and assessment of current
                   practice},
  volume =        {21},
  year =          {2011},
  issn =          {0959-1524},
}

@article{camara:quelhos:pinto:2016,
  author =        {C{\^a}mara, Maur{\'\i}cio M and Quelhas, Andr{\'e} D and
                   Pinto, Jos{\'e} Carlos},
  journal =       {Processes},
  number =        {4},
  pages =         {44},
  publisher =     {MDPI},
  title =         {Performance evaluation of real industrial {RTO}
                   systems},
  volume =        {4},
  year =          {2016},
}

@article{cutler:perry:lu:1983,
  author =        {Cutler, C. R. and Perry, R. T.},
  journal =       {Comput. Chem. Eng.},
  pages =         {663-667},
  title =         {Real time optimization with multivariable control is
                   required to maximize profits},
  volume =        {7},
  year =          {1983},
}

@book{seborg:edgar:mellichamp:doyle:2017,
  address =       {New York},
  author =        {Dale E. Seborg and Thomas F. Edgar and
                   Duncan A. Mellichamp and Francis J. Doyle},
  edition =       {fourth},
  publisher =     {John Wiley and Sons},
  title =         {Process Dynamics and Control},
  year =          {2017},
}

@article{crowe:1996,
  author =        {Crowe, Cameron M},
  journal =       {J. Proc. Cont.},
  number =        {2-3},
  pages =         {89--98},
  publisher =     {Elsevier},
  title =         {Data reconciliation—progress and challenges},
  volume =        {6},
  year =          {1996},
}

@article{roberts:williams:1981,
  author =        {Roberts, P. D. and Williams, T. W. C.},
  journal =       {Automatica},
  number =        {1},
  pages =         {199--209},
  title =         {On an algorithm for combined system optimisation and
                   parameter estimation},
  volume =        {17},
  year =          {1981},
}

@article{krishnamoorthy:skogestad:2018,
  author =        {Krishnamoorthy, Dinesh and Foss, Bjarne and
                   Skogestad, Sigurd},
  journal =       {Comput. Chem. Eng.},
  pages =         {34--45},
  publisher =     {Elsevier},
  title =         {Steady-state real-time optimization using transient
                   measurements},
  volume =        {115},
  year =          {2018},
}

@inproceedings{rawlings:angeli:bates:2012,
  address =       {Maui, HI},
  author =        {James B. Rawlings and David Angeli and Cuyler Bates},
  booktitle =     {{IEEE} Conference on Decision and Control ({CDC})},
  month =         {December},
  pages =         {3851-3861},
  title =         {Fundamentals of Economic Model Predictive Control},
  year =          {2012},
}

@article{box:1957,
  author =        {George E. P. Box},
  journal =       {Journal of the Royal Statistical Society Series C},
  month =         {June},
  number =        {2},
  pages =         {81-101},
  title =         {Evolutionary Operation: A Method for Increasing
                   Industrial Productivity},
  volume =        {6},
  year =          {1957},
  doi =           {10.2307/2985505},
  url =           {https://ideas.repec.org/a/bla/jorssc/v6y1957i2p81-101.html},
}

@article{skogestad:2000b,
  author =        {Skogestad, Sigurd},
  journal =       {Comput. Chem. Eng.},
  number =        {2-7},
  pages =         {569--575},
  publisher =     {Elsevier},
  title =         {Self-optimizing control: The missing link between
                   steady-state optimization and control},
  volume =        {24},
  year =          {2000},
}

@article{krishnamoorthy:skogestad:2022,
  author =        {Krishnamoorthy, Dinesh and Skogestad, Sigurd},
  journal =       {Comput. Chem. Eng.},
  pages =         {107723},
  publisher =     {Elsevier},
  title =         {Real-time optimization as a feedback control
                   problem--a review},
  volume =        {161},
  year =          {2022},
}

@article{bonvin:pannocchia:2024,
  author =        {Bonvin, Dominique and Pannocchia, Gabriele},
  journal =       {Comput. Chem. Eng.},
  pages =         {108839},
  publisher =     {Elsevier},
  title =         {On speeding-up modifier-adaptation schemes for
                   real-time optimization},
  volume =        {191},
  year =          {2024},
}

@article{chachuat:srinivasan:bonvin:2009,
  author =        {B. Chachuat and B. Srinivasan and D. Bonvin},
  journal =       {Comput. Chem. Eng.},
  number =        {10},
  pages =         {1557 - 1567},
  title =         {Adaptation strategies for real-time optimization},
  volume =        {33},
  year =          {2009},
}

@article{papasavvas:ferreira:marchetti:bonvin:2019,
  author =        {Papasavvas, Aris and de Avila Ferreira, Tafarel and
                   Marchetti, Alejandro G and Bonvin, Dominique},
  journal =       {Comput. Chem. Eng.},
  pages =         {285--293},
  publisher =     {Elsevier},
  title =         {Analysis of output modifier adaptation for real-time
                   optimization},
  volume =        {121},
  year =          {2019},
}

@article{bindlish:2025,
  author =        {Bindlish, Rahul},
  journal =       {Ind. Eng. Chem. Res.},
  number =        {50},
  pages =         {23810--23823},
  publisher =     {ACS Publications},
  title =         {{An Industrial Perspective on Applying Process
                   Knowledge and Mathematical Programming for RTO, MPC,
                   NMPC, and Power Scheduling}},
  volume =        {64},
  year =          {2025},
}

@article{forbes:marlin:macgregor:1994,
  author =        {Forbes, J. F. and Marlin, T. E. and MacGregor, J. F.},
  journal =       {Comput. Chem. Eng.},
  number =        {6},
  pages =         {497--510},
  title =         {Model adequacy requirements for optimizing plant
                   operations},
  volume =        {18},
  year =          {1994},
  issn =          {0098-1354},
}

@article{hornik:stinchcombe:white:1989,
  author =        {Hornik, Kurt and Stinchcombe, Maxwell and
                   White, Halbert},
  journal =       {Neural Netw.},
  number =        {5},
  pages =         {359--366},
  publisher =     {Elsevier},
  title =         {Multilayer feedforward networks are universal
                   approximators},
  volume =        {2},
  year =          {1989},
}

@article{chen:rubanova:bettencourt:duvenaud:2018,
  author =        {Chen, Ricky TQ and Rubanova, Yulia and
                   Bettencourt, Jesse and Duvenaud, David K},
  journal =       {Adv. Neural Inf. Process. Syst.},
  pages =         {6571--6583},
  title =         {Neural ordinary differential equations},
  volume =        {31},
  year =          {2018},
}

@article{sansana:joswiak:castillo:wang:rendall:chiang:reis:2021,
  author =        {Sansana, Joel and Joswiak, Mark N and Castillo, Ivan and
                   Wang, Zhenyu and Rendall, Ricardo and Chiang, Leo H and
                   Reis, Marco S},
  journal =       {Comput. Chem. Eng.},
  pages =         {107365},
  publisher =     {Elsevier},
  title =         {Recent trends on hybrid modeling for Industry 4.0},
  volume =        {151},
  year =          {2021},
}

@article{schweidtmann:zhang:stosch:2024,
  author =        {Schweidtmann, Artur M and Zhang, Dongda and
                   Von Stosch, Moritz},
  journal =       {Digital Chemical Engineering},
  pages =         {100136},
  publisher =     {Elsevier},
  title =         {A review and perspective on hybrid modeling
                   methodologies},
  volume =        {10},
  year =          {2024},
}

@article{mukherjee:zavala:2026,
  author =        {Mukherjee, Angan and Zavala, Victor M},
  journal =       {Curr. Opin. Chem. Eng.},
  pages =         {101228},
  publisher =     {Elsevier},
  title =         {Physics-constrained machine learning for chemical
                   engineering},
  volume =        {51},
  year =          {2026},
}

@article{raissi:perdikaris:karniadakis:2019,
  author =        {Raissi, Maziar and Perdikaris, Paris and
                   Karniadakis, George E},
  journal =       {J. Comput. Phys.},
  pages =         {686--707},
  publisher =     {Elsevier},
  title =         {Physics-informed neural networks: A deep learning
                   framework for solving forward and inverse problems
                   involving nonlinear partial differential equations},
  volume =        {378},
  year =          {2019},
}

@article{sholokhov:liu:mansour:nabi:2023,
  author =        {Sholokhov, Aleksei and Liu, Yuying and
                   Mansour, Hassan and Nabi, Saleh},
  journal =       {Sci. Rep.},
  number =        {1},
  pages =         {10166},
  publisher =     {Nature Publishing Group UK London},
  title =         {Physics-informed neural ODE (PINODE): embedding
                   physics into models using collocation points},
  volume =        {13},
  year =          {2023},
}

@misc{golder:roy:hasan:2025,
  author =        {Golder, Rahul and Roy, Bimol Nath and Hasan, MM},
  journal =       {arXiv preprint arXiv:2512.05881},
  title =         {DAE-HardNet: A Physics Constrained Neural Network
                   Enforcing Differential-Algebraic Hard Constraints},
  year =          {2025},
}

@inproceedings{constante:chen:li:2026,
  author =        {Constante, Gonzalo E and Chen, Hao and Li, Can},
  journal =       {Advances in Neural Information Processing Systems},
  pages =         {105075--105100},
  title =         {Enforcing Hard Linear Constraints in Deep Learning
                   Models with Decision Rules},
  volume =        {38},
  year =          {2026},
}

@article{lueg:alves:schicksnus:kitchin:laird:biegler:2025,
  author =        {Lueg, Laurens R and Alves, Victor and
                   Schicksnus, Daniel and Kitchin, John R and
                   Laird, Carl D and Biegler, Lorenz T},
  journal =       {arXiv preprint arXiv:2504.04665},
  title =         {A Simultaneous Approach for Training Neural
                   Differential-Algebraic Systems of Equations},
  year =          {2025},
}

@article{kumar:rawlings:2023a,
  author =        {Kumar, Pratyush and Rawlings, James B.},
  journal =       {Comput. Chem. Eng.},
  pages =         {108314},
  title =         {Structured nonlinear process modeling using neural
                   networks and application to economic optimization},
  volume =        {177},
  year =          {2023},
  doi =           {https://doi.org/10.1016/j.compchemeng.2023.108314},
}

@article{thompson:connors:zavala:venturelli:2026,
  author =        {Thompson, Jaron and Connors, Bryce M and
                   Zavala, Victor M and Venturelli, Ophelia S},
  journal =       {Proc. Natl. Acad. Sci. USA},
  number =        {13},
  pages =         {e2517661123},
  publisher =     {National Academy of Sciences},
  title =         {Physics-constrained neural ordinary differential
                   equation models to discover and predict microbial
                   community dynamics},
  volume =        {123},
  year =          {2026},
}

@article{pantelides:baldea:georgiou:gopaluni:mehmet:sheth:zavala:georgakis:2025,
  author =        {Pantelides, Constantinos and Baldea, Michael and
                   Georgiou, Apostolos T and Gopaluni, Bhushan and
                   Mehmet, Mercang{\"o}z and Sheth, Kiran and
                   Zavala, Victor M and Georgakis, Christos},
  journal =       {Comput. Chem. Eng.},
  title =         {{From Automated to Autonomous Process Operations}},
  volume =        {196},
  year =          {2025},
  doi =           {https://doi.org/10.1016/j.compchemeng.2025.109064},
}

@article{luyben:tyreus:1998,
  author =        {Luyben, Michael L and Tyr{\'e}us, Bj{\"o}rn D},
  journal =       {Comput. Chem. Eng.},
  number =        {7-8},
  pages =         {867--877},
  publisher =     {Elsevier},
  title =         {An industrial design/control study for the vinyl
                   acetate monomer process},
  volume =        {22},
  year =          {1998},
}

@article{chen:dave:mcavoy:luyben:2003,
  author =        {Chen, Rong and Dave, Kedar and McAvoy, Thomas J and
                   Luyben, Michael},
  journal =       {Ind. Eng. Chem. Res.},
  number =        {20},
  pages =         {4478--4487},
  publisher =     {ACS Publications},
  title =         {A nonlinear dynamic model of a vinyl acetate process},
  volume =        {42},
  year =          {2003},
}

@article{ward:mellichamp:doherty:2004,
  author =        {Ward, Jeffrey D and Mellichamp, Duncan A and
                   Doherty, Michael F},
  journal =       {Ind. Eng. Chem. Res.},
  number =        {14},
  pages =         {3957--3971},
  publisher =     {ACS Publications},
  title =         {Importance of process chemistry in selecting the
                   operating policy for plants with recycle},
  volume =        {43},
  year =          {2004},
}

@book{rawlings:ekerdt:2020,
  address =       {Santa Barbara, CA},
  author =        {James B. Rawlings and John G. Ekerdt},
  edition =       {2nd, Paperback},
  note =          {664 pages, ISBN 978-0-9759377-4-7},
  publisher =     {Nob Hill Publishing},
  title =         {Chemical Reactor Analysis and Design Fundamentals},
  year =          {2020},
}

@article{wachter:biegler:2006,
  author =        {W{\"a}chter, A. and Biegler, L. T.},
  journal =       {Math. Prog.},
  number =        {1},
  pages =         {25--57},
  publisher =     {Springer},
  title =         {{On the implementation of a primal-dual
                   interior-point filter line-search algorithm for
                   large-scale nonlinear programming}},
  volume =        {106},
  year =          {2006},
}

@article{andersson:gillis:horn:rawlings:diehl:2019,
  author =        {Joel A. E. Andersson and Joris Gillis and Greg Horn and
                   James B. Rawlings and Moritz Diehl},
  journal =       {Math. Prog. Comp.},
  month =         {Mar},
  number =        {1},
  pages =         {1-36},
  title =         {{CasADi}---A software framework for nonlinear
                   optimization and optimal control},
  volume =        {11},
  year =          {2019},
  doi =           {10.1007/s12532-018-0139-4},
}

@misc{bradbury:frostig:hawkins:johnson:et_al:2018,
  author =        {James Bradbury and Roy Frostig and Peter Hawkins and
                   Matthew James Johnson and Chris Leary and
                   Dougal Maclaurin and George Necula and Adam Paszke and
                   Jake Vander{P}las and Skye Wanderman-{M}ilne and
                   Qiao Zhang},
  title =         {{JAX}: composable transformations of
                   {P}ython+{N}um{P}y programs},
  year =          {2018},
  url =           {http://github.com/jax-ml/jax},
}

@phdthesis{kidger:2021,
  author =        {Patrick Kidger},
  school =        {University of Oxford},
  title =         {{O}n {N}eural {D}ifferential {E}quations},
  year =          {2021},
}

@software{flax:2020,
  author =        {Jonathan Heek and Anselm Levskaya and Avital Oliver and
                   Marvin Ritter and Bertrand Rondepierre and
                   Andreas Steiner and Marc van {Z}ee},
  title =         {{F}lax: A neural network library and ecosystem for
                   {JAX}},
  year =          {2024},
  url =           {http://github.com/google/flax},
}

@article{kingma:ba:2014,
  author =        {Kingma, Diederik P and Ba, Jimmy},
  journal =       {arXiv preprint arXiv:1412.6980},
  title =         {Adam: A method for stochastic optimization},
  year =          {2014},
}

@article{liu:nocedal:1989,
  author =        {Liu, Dong C and Nocedal, Jorge},
  journal =       {Math. Prog.},
  number =        {1},
  pages =         {503--528},
  publisher =     {Springer},
  title =         {On the limited memory BFGS method for large scale
                   optimization},
  volume =        {45},
  year =          {1989},
}

@article{fronk:petzold:2025,
  author =        {Fronk, Colby and Petzold, Linda},
  journal =       {Chaos},
  number =        {11},
  publisher =     {AIP Publishing},
  title =         {The vanishing gradient problem for stiff neural
                   differential equations},
  volume =        {35},
  year =          {2025},
}

@misc{peng:liq:boy:2025,
  author =        {Peng, Wenqing and Liu, Zhi-Song and Boy, Michael},
  journal =       {arXiv preprint arXiv:2505.05625},
  title =         {Spin-ode: Stiff physics-informed neural ode for
                   chemical reaction rate estimation},
  year =          {2025},
}

@incollection{lecun:leon:orr:muller:2002,
  author =        {LeCun, Yann and Bottou, L{\'e}on and Orr, Genevieve B and
                   M{\"u}ller, Klaus-Robert},
  booktitle =     {Neural networks: Tricks of the trade},
  pages =         {9--50},
  publisher =     {Springer},
  title =         {Efficient backprop},
  year =          {2002},
}

@article{dake:ilagan:banerjee:scott:rawlings:2024,
  author =        {Dake, Prithvi and Ilagan, Maria Rikaela and
                   Banerjee, Shoili and Scott, Susannah L and
                   Rawlings, James B},
  journal =       {J. Vac. Sci. Technol. A.},
  month =         {Sep},
  number =        {6},
  publisher =     {American Vacuum Society},
  title =         {Identifying kinetic models from reactor measurements},
  volume =        {42},
  year =          {2024},
  doi =           {10.1116/6.0003846},
  issn =          {1520-8559},
}

@dataset{dake:bindlish:rawlings:2026a,
  author =        {Dake, Prithvi and Bindlish, Rahul and
                   Rawlings, James},
  month =         {Jul},
  publisher =     {Zenodo},
  title =         {Dataset for `{A} tale of perfect fit and phantom
                   optima: how data-driven models can fail in real- time
                   optimization'},
  year =          {2026},
  doi =           {10.5281/zenodo.21464342},
  url =           {https://doi.org/10.5281/zenodo.21464342},
}
